\documentclass[a4paper,fleqn]{cas-dc}

\usepackage[authoryear]{natbib}

\usepackage{tabularx} %
\usepackage{booktabs} %
\usepackage{rotating} %
\usepackage{longtable} %
\usepackage{forest} %
\usepackage{tikz} %
\usetikzlibrary{shapes.geometric, arrows, positioning, calc} %
\usepackage{booktabs}
\usepackage{array}
\usepackage{tabularx}
\usepackage[labelfont=bf,format=plain,labelsep=colon,textformat=period,font={small,singlespacing},justification=justified,singlelinecheck=false,skip=6pt,belowskip=-6pt]{caption}
\usepackage{tcolorbox}  

\def\tsc#1{\csdef{#1}{\textsc{\lowercase{#1}}\xspace}}
\tsc{AI}
\tsc{LLMs}

\begin{document}
\let\WriteBookmarks\relax %
\def\floatpagepagefraction{1} %
\def\textpagefraction{.001} %

\shorttitle{AI Safety in Generative AI Large Language Models: A Survey}

\shortauthors{J. Chua, Y. Li, S. Yang, C. Wang and L. Yao}

\title [mode = title]{AI Safety in Generative AI Large Language Models: A Survey}

\author[1,2]{Jaymari Chua}

\cormark[1]

\ead{jace.chua@csiro.au}

\credit{Ch.1,5-8,9}

\affiliation[1]{organization={CSIRO's Data61},
        addressline={Australian Technology Park},
        addressline={13 Garden Street},
        city={Eveleigh},
        postcode={2015},
        state={NSW},
        country={Australia}}

\affiliation[2]{organization={University of New South Wales},
        addressline={UNSW Sydney},
        city={Kensington},
        postcode={2052},
        state={NSW},
        country={Australia}}

\author[1]{Yun Li}

\ead{yun.li@csiro.au}

\credit{Ch.2-4}

\author[1,2]{Shiyi Yang}

\ead{shiyi.yang@csiro.au}

\credit{}

\author[1]{Chen Wang}

\ead{chen.wang@csiro.au}

\author[1,2]{Lina Yao}

\ead{lina.yao@csiro.au}

\cortext[cor1]{Corresponding author}
\cortext[cor2]{Principal corresponding author}

\begin{abstract}
Large Language Model (LLMs) such as ChatGPT that exhibit generative AI capabilities are facing accelerated adoption and innovation. The increased presence of Generative AI (GAI) inevitably raises concerns about the risks and safety associated with these models. This article provides an up-to-date survey of recent trends in AI safety research of GAI-LLMs from a computer scientist's perspective: specific and technical. In this survey, we explore the background and motivation for the identified harms and risks in the context of LLMs being generative language models; our survey differentiates by emphasising the need for unified theories of the distinct safety challenges in the research development and applications of LLMs. We start our discussion with a concise introduction to the workings of LLMs, supported by relevant literature. Then we discuss earlier research that has pointed out the fundamental constraints of generative models, or lack of understanding thereof (e.g., performance and safety trade-offs as LLMs scale in number of parameters). We provide a sufficient coverage of LLM alignment -- delving into various approaches, contending methods and present challenges associated with aligning LLMs with human preferences. By highlighting the gaps in the literature and possible implementation oversights, our aim is to create a comprehensive analysis that provides insights for addressing AI safety in LLMs and encourages the development of aligned and secure models. We conclude our survey by discussing future directions of LLMs for AI safety, offering insights into ongoing research in this critical area.
\end{abstract}

\begin{highlights}
\item Up-to-date.
\item Computer scientists' perspective: specific and technical.
\end{highlights}


\maketitle %

\section{Introduction}
Large Language Models (LLMs), often celebrated as revolutionary Generative AI (GAI) models with their astounding generative prowess, present several AI Safety challenges. This article provides a comprehensive review of the AI safety risks inherent to LLMs as documented in the extant research literature. A significant challenge lies in the lack of a dedicated survey paper in AI Safety that aptly tackles observed issues with GAI-LLMs as inherited from technical composition of what makes an LLMs, i.e., its data pre-training, model architecture, and prompt usage. Current research often tends to be overly broad, delving into topics such as X-risk, AI ethics boundaries and government policy changes, which, while important, do not directly focus on the functional aspects of GAI-LLMs. In this work, we examine the known limitations of generative models as documented in the prior literature and further identify and discuss new safety challenges that emerge as a consequence of the unaddressed technical aspects that make an LLMs. After pinpointing the challenges related to safety in generative AI and large language models, we have identified a variety of key studies and topics: bias and fairness, toxicity and harmful content generation, misinformation, privacy and security concerns, adversarial attacks, infringement of intellectual property, ethics and human preference alignment and safety at scale. This article also reviews promising avenues for subsequent GAI-LLMs safety research, emphasizing the breadth of vital elements that must be taken into account by both LLM practitioners and application developers as GAI-LLMs become more capable as agents.

\subsection{Strategy for Literature Search}
This work surveys AI Safety in the context of Generative AI and LLMs from computer science publications that has a focus on AI and NLP: ACL, NAACL, EACL, EMNLP, CoNLL, COLING, SIGIR, IJCAI, AAAI, ICML, ICDM, KDD, and NeurIPS. Relevant keywords such as AI Safety, Generative AI, and Large Language Models are used to search for relevant papers. Related work appearing in the found papers is also included in the scope of this survey, which may include technical articles from technology companies that are researching AI Safety, e.g., Anthropic, Google DeepMind, Meta, Microsoft Research, and OpenAI. The domain of AI Safety and Generative AI research is extensive and complex, with input coming from multiple sectors such as academia, industry, and government; building on the central literature previously mentioned, it is important to examine additional sources that provide varied viewpoints and insights. These additional sources enhance our comprehension of the field and underscore the collaborative endeavors needed to tackle the issues associated with advanced AI systems. The forthcoming categories highlight pivotal areas where substantial computer science research and policymaking are underway:

\begin{itemize}
    \item \textit{Taxonomies of Risks and Harms of Large Language Models.} \cite{weidinger2024holistic} underscore the significance of research aimed at identifying and categorizing the safety risks linked to LLMs, especially those with potential sociotechnical impacts. This type of work, as demonstrated by \cite{bommasani2022opportunities}, \cite{shelby2023sociotechnical}, and \cite{weidinger2022taxonomy}, offers crucial taxonomies and frameworks for comprehending the potential dangers of GAI-LLMs. \cite{liu2024trustworthy} provide an in-depth analysis of LLMs alignment risks, systematically organizing them by topic and discussing recent advancements in evaluation research. Collectively, these survey papers add to a growing body of literature focused on identifying and understanding the potential adverse effects of LLMs. \cite{bird2023typology} examine the harms of text-to-image generative models, such as bias in the generated content, producing inappropriate content, and violating copyright. While this survey paper concentrates on the technical aspects of large language models, a thorough evaluation of their societal impacts, including potential harms, is explored in \cite{shevlane2023model} and \cite{solaiman2023evaluating}, building on their prior work in \cite{solaiman2019release}. The foundational work "Concrete Problems in AI Safety" \cite{amodei2016concrete} has greatly influenced discussions on advanced AI, while Google DeepMind's evolving taxonomy \cite{morris2023levels, DeepMind_Introducing_the_Frontier_Safety_Framework} has aided in identifying research that establishes this field. As \cite{weidinger2024holistic} emphasize, the continuous development of such taxonomies is vital for fostering a shared understanding among researchers, policymakers, and other stakeholders during this critical period for this emerging technology.

\item \textit{Government/Policymaking Conferences.} Beyond the contributions of the research community, the participation of government policymakers in collaborative initiatives. These have been instrumental in the generation of quality reports on AI safety \cite{White_House}. For example, both Google DeepMind's evaluation approach \cite{liang2022holistic} and Anthropic's responsible scaling policy \cite{AnthropicResponsibleScalingPolicy} cites the \cite{White_House}. These reports aid in cultivating a comprehensive comprehension of the intricate landscape enveloping AI systems \cite{international_scientific_report_on_the_safety_of_advanced_ai}. As can be expected, papers from these conferences in the field tend to centre the discussion on governing AI safety risks, such as AI trustworthiness, monitoring, fairness, and privacy.

\item \textit{Corporate AI Safety Boards and AI Safety Statements.} OpenAI publish their approach to AI Safety through their blog with ongoing updates and changes with their OpenAI Safety Board \cite{OpenAI_Board_Forms_Safety_and_Security_Committee}, AISafety Update \cite{OpenAI_Safety_Update}, and OpenAI Safety Standards \cite{OpenAI_Safety_Standards}. Google DeepMind. \cite{weidinger2024holistic} mention GoogleDeepMind's AI safety board and also cite a few more links or resources to their governing AI safety principles \cite{Google_AI_Principles_2023_Progress_Update}. Anthropic's contribution to the discourse on evaluating AI systems offers a more pragmatic and focused approach. \cite{ganguli2023challenges} highlight the inherent complexities associated with specific evaluations available for GAI-LLMs, such as the Measuring Massive Multitask Evaluation (MMLU) benchmark \cite{hendrycks2021measuring} and the Bias Benchmark for Question Answering (BBQ) \cite{parrish2022bbq}, alongside broader assessments like Holistic Evaluation of Language Models (HELM) \cite{liang2023holistic} and BigBench \cite{srivastava2023imitation}. The challenge with creating such comprehensive evaluations underscore the massive effort necessary to create them.

\item \textit{AI Safety Evaluations.} Comprehensive evaluation frameworks such as HELM \cite{liang2023holistic} and BigBench \cite{srivastava2023imitation} offer streamlined APIs for benchmarking LLMs against diverse tasks. These frameworks incorporate benchmarks from seminal research, including TruthfulQA \cite{lin2021truthfulqa}, thereby providing a continually evolving snapshot of the field's progress in addressing established risks and harms of AI safety and as agents \cite{ye2024miraievaluatingllmagents, zhang2024task}. Through hosting on open-source code repositories on platforms like GitHub and HuggingFace, the researchers facilitate ongoing community contributions and updates.
\end{itemize}

\subsection{Comparisons with Other Surveys}
This survey paper sets itself apart from other current surveys on large language models and AI safety by systematically investigates safety issues in LLMs through a high-level categorization of training data, model training, prompting, alignment, and scaling. By breaking down the discussion of advanced applications such as ChatGPT, Gemini, and others into distinct aspects, we can narrow safety concerns at their technical scope and understand downstream problems. Our reductionist approach is in line with recent efforts, e.g. Guaranteed Safe AI by \cite{dalrymple2024guaranteed}, to synthesize and categorize the risks associated with GAI-LLMs and as future agents. For example, \cite{gabriel2024ethics} offer an in-depth examination of ethical challenges in advanced AI assistants, focusing on value alignment, safety, and potential misuse. Similarly, \cite{bommasani2022opportunities} provide a broad view of the opportunities and risks presented by foundation models, including LLMs. \cite{liu2024trustworthy} add to this discussion by identifying known risks in generative AI systems and proposing frameworks for enhancing their trustworthiness. Whereas other survey work provide essential insights, our survey distinguishes itself by systematically correlating these identified risks with particular methodologies intrinsic to LLM architectures, specifically in-context learning, prompting, and reinforcement learning. This approach allows us to identify the technical origins of safety issues and trace their propagation through the system, offering a more detailed understanding of how and where interventions might be most effective. We also incorporate recent developments in the evaluation of deployed generative AI systems, as demonstrated by \cite{weidinger2024holistic}. Their work, which shares lessons learned from the development of Google's Gemini model and insights into DeepMind's internal governance structures, provides a crucial perspective on the practical challenges of implementing safety measures in real-world AI systems. To further distinguish our survey, we include an analysis of recent work by \cite{zhao2023survey}, who provide a comprehensive overview of evaluation methods for LLMs. Their taxonomy of evaluation metrics and datasets offers valuable context for our component-based analysis, helping to bridge the gap between theoretical safety concerns and practical assessment methodologies. By synthesizing these diverse perspectives and organizing them around a component-based framework, our survey makes a unique contribution to the field. It not only consolidates existing knowledge on LLM safety but also provides a structured approach for identifying and addressing safety issues at their source. This approach is particularly valuable for researchers and practitioners aiming to improve the robustness and reliability of LLM-based systems across various applications and domains.

\begin{table*}[htb]
\centering
\caption{Taxonomy of Safety Concerns and Challenges in Large Language Models}
\label{tab:taxonomy}
\small
\begin{tabularx}{\linewidth}{>{\raggedright\arraybackslash}p{0.19\linewidth}p{0.22\linewidth}X}
\toprule
\textbf{Category} & \textbf{Subcategories} & \textbf{Descriptions} \\
\midrule
Data Safety (Sec~\ref{sec:data_safety}) & 
    Toxicity & Generating Toxic and Harmful Content (Sec~\ref{sec:toxicity}) \\
    & Bias & Generating Biased Answers (Sec~\ref{sec:bias}) \\
    & Data Privacy & Leaking Private Information from Training Data (Sec~\ref{sec:data_privacy}) \\
    & Copyright & Inadvertently Breaching Copyright with Generation (Sec~\ref{sec:copyright}) \\
\midrule
Model Safety (Sec~\ref{sec:model_safety}) & 
    Misinformation & Generating Wrong Content with Hallucination (Sec~\ref{sec:misinformation}) \\
    & Eval Problem & Insufficient Evaluation and Verification Protocols (Sec~\ref{sec:eval_problem}) \\
    & Explainability & Lack of Transparency and Explainability (Sec~\ref{sec:explainability}) \\
    & Inference Privacy & Sensitive Information and Unintended Patent-busting (Sec~\ref{sec:inference_privacy}) \\
\midrule
Prompt Safety (Sec~\ref{sec:prompt_safety}) & 
    Prompting & Attacking with Prompt Injection and Jailbreaking (Sec~\ref{sec:prompt_injection_and_jailbreaking}) \\
    & Guardrails & Defending with Guardrails (Sec~\ref{sec:guardrails}) \\
\midrule
Alignment (Sec~\ref{sec:alignment}) & 
    Philosophy of Alignment & Ethics and Morality Alignment (Sec~\ref{sec:ethics}) \\
    & Agentic Alignment & Treating AI agents differently in AI safety contexts (Sec~\ref{sec:agentic_alignment}) \\
    & Moral Values & Defining Value Alignment for Large Language Models (Sec~\ref{sec:defining_value}) \\
    & Humane & Generating Content that Aligns with Human (Sec~\ref{sec:generating_content}) \\
    & Alignment Techniques & Methods to Align LLMs (Sec~\ref{sec:methods_to_align}) \\
\midrule
Safety at Scale (Sec~\ref{sec:safety_at_scale}) & 
    Scalable Oversight & Scaling GAI-LLMs Beyond Human Supervision (Sec~\ref{sec:scalable_oversight}) \\
    & Emergent Abilities & Increasingly Capable GAI-LLMs in Reasoning and Tasks (Sec~\ref{sec:emergent_abilities}) \\
    & Knowledge Distillation & KD Techniques for Model Alignment and Efficiency (Sec~\ref{sec:knowledge_distillation}) \\
    & Catastrophic Forgetting & GAI-LLMs Memory Retention challenges Performance (Sec~\ref{sec:catastrophic_forgetting}) \\
\bottomrule
\end{tabularx}
\end{table*}

\subsection{The Main Contributions of the Survey}

In this survey, our main contributions are summarized as follows: 
\begin{itemize}
\item We provide a systematic investigation of safety issues in LLMs through a novel component-based framework, categorizing concerns across training data, model training, prompting, alignment, and scaling.
\item We correlate identified risks with specific LLM methodologies, particularly in-context learning, prompting, and reinforcement learning, enabling a more precise understanding of the technical origins of safety issues.
\item We integrate a comprehensive analysis of LLM prompting and alignment techniques to human preferences, bridging the gap between theoretical safety concerns and practical assessment methodologies.
\item We position the conversation on model alignment within the extensive body of AI safety literature, exploring different philosophical views on language models and how they are uniquely treated compared to AI agent safety. According to our research, we differentiate reinforcement learning strategies where agents, regardless of being designed as language assistants, can embed safety measures concerning humans in any mutual setting.
\item Through this reductionist method, we bring together various viewpoints from current literature, presenting a unique and organized framework to pinpoint and tackle LLM safety concerns at their origin. This approach provides essential insights into the most effective intervention points, offering a vital perspective for researchers and practitioners focused on increasing the safety of LLM-based systems with increasing parameter scales and consequential emergent capabilities.
\end{itemize}

\subsection{The Outline of the Survey}

In this work, a brief background of LLMs is provided in Sec~\ref{background}. The paper then presents a comprehensive taxonomy of safety concerns and challenges associated with large language models, organized into five main categories, as elaborately illustrated in Table \ref{tab:taxonomy}. Data Safety in Sec~\ref{sec:data_safety} addresses issues such as toxicity, bias, data privacy, and copyright infringement. Model Safety in Sec~\ref{sec:model_safety} explores misinformation, evaluation problems, explainability challenges, and inference privacy concerns. Prompt Safety or Usage Safety in Sec~\ref{sec:prompt_safety} focuses on potential attacks like prompt injection and jailbreaking, as well as defensive strategies. Alignment or Ethics Safety in Sec~\ref{sec:alignment} delves into philosophical aspects of alignment, agnostic alignment approaches, value alignment for large language models, and methods for aligning LLMs with human values and ethical principles. Finally, Safety at Scale in Sec~\ref{sec:safety_at_scale} examines scalable oversight mechanisms, emergent abilities, knowledge distillation techniques, and the risk of catastrophic forgetting in large-scale models. This structured approach provides a thorough overview of the safety landscape in GAI-LLMs research. We finally give some research trends that are worth investigating in the future in Sec~\ref{future_work} and this survey is concluded in Sec~\ref{conclusion}.

\section{Background} \label{background}
Large language models have revolutionized natural language processing, building upon the transformer architecture introduced by Vaswani et al. \cite{vaswani2017attention}. LLMs can be categorized into three main architectural paradigms: encoder-decoder (e.g., BART \cite{lewis2019bart}, T5 \cite{raffel2020exploring}), encoder-only (e.g., BERT \cite{devlin2018bert}), and decoder-only (e.g., GPT \cite{brown2020language}) models, each optimized for specific tasks such as language understanding or text generation. A key innovation in LLMs is In-Context Learning (ICL), enabling models to adapt to new tasks without explicit fine-tuning, thus reducing computational overhead and enhancing versatility \cite{dong2023survey}. Recent research has focused on optimizing ICL performance through strategies like meta-training frameworks, symbol tuning \cite{wei2023symbol}, and self-supervised training data development. The ongoing evolution of LLMs continues to advance the field, pushing the boundaries of natural language understanding and generation towards more sophisticated AI systems. For a technical review of LLMs, see \cite{zhao2023survey}. And for a technical review of Visual Language models, see \cite{bordes2024introduction}.

\subsection{Model Architecture}
The success of large language models begins with the proposal of the transformer model \cite{vaswani2017attention},  which is the foundation for subsequent LLMs due to its remarkable parallelizability and capacity. Thus, we first introduce the preliminary information about the transformer and its key components and then introduce the mainstream architectures of LLMs derived from the vanilla transformer.

\subsubsection{Preliminary: Transformer.}
The vanilla transformer \cite{vaswani2017attention} adopts an encoder-decoder structure that leverages stacked multi-head self-attention and position-wise feed-forward modules as the building blocks. Within the blocks, residual connections enable seamless information flow, and normalization layers ensure output stability in each sub-layer. 
Of these blocks, multi-head attention is a critical component,  allowing the model to attend to representations from different sub-spaces jointly and in parallel. It can be mathematically formulated as follows:
\begin{gather}
   \text{MultiHead}(Q,K,V) = \text{Concat}(\text{h}_1,\ldots,\text{h}_h)W^O, \\
   \text{where, } {h}_i = \text{Attention}(QW^Q_i,KW^K_i,VW^V_i), \\
   \text{Attention}(Q,K,V) = \text{softmax}\left(\frac{QK^T}{\sqrt{d_k}}V\right),
\end{gather} where $W^Q_i\in\mathbb{R}^{d_{\text{model}}\times d_k}$, $W^K_i\in\mathbb{R}^{d_{\text{model}}\times d_k}$, and $W^V_i\in\mathbb{R}^{d_{\text{model}}\times d_v}$ are matrices to project input embeddings of dimension $d$ into query, key and value matrices of dimensions $d_k$, $d_k$, and $d_v$ respectively; and $W^O\in\mathbb{R}^{hd_v\times d_{\text{model}}}$ are output projections.

Apart from multi-head attention identical to those in the encoder, the cross-attention modules are further introduced to the decoder with queries from decoder layers while keys and values from encoder layers. This allows positions in the decoder to attend to all positions in the input sequences.

Subsequent studies optimize the attention module to enhance computational efficiency and model effectiveness. For instance, Qiu et al. \cite{qiu2019blockwise} propose Blockwise Attention, a lightweight variant that employs sparse blocks to reduce computational costs during training and inferencing. Ainslie et al. \cite{ainslie-etal-2020-etc} and Zaheer et al.~\cite{zaheer2020big} combine global and local attention mechanisms to handle longer contexts. Wang et al. \cite{wang2020linformer} present an approximation of the self-attention mechanism using a low-rank matrix, leading to more computationally efficient models with linear time and space complexity.

Another critical component of the transformer is positional encoding. As the self-attention operation is permutation invariant and the transformer lacks recurrent or convolutional modules, positional information must be provided explicitly to reveal the order of input sequences.  The vanilla transformer adopts sinusoidal positional encoding to represent position information as follows:

\begin{gather}
   PE_{(pos,2i)} = \sin\left(\frac{pos}{10000^{2i/d_{\text{model}}}}\right), \\
   PE_{(pos,2i+1)} = \cos\left(\frac{pos}{10000^{2i/d_{\text{model}}}}\right).
\end{gather}

With position information, leftward information flow in the decoder self-attention layer, i.e., positions attending to the future, is prevented by masking out values corresponding to illegal connections, thus preserving the auto-regressive property. Recent iterations of transformers have introduced enhanced positional embedding methods. For example, Shaw et al. \cite{shaw2018self} and Dai et al. \cite{dai2019transformer}  explore relative positional encoding, incorporating relative position dependencies. Su et al. \cite{su2021roformer} introduce RoFormer, which employs rotation matrices to encode absolute positions and inject relative positional information into self-attention layers.

\begin{table*}[ht*]
\caption{Mainstream architectures of LLMs.}\label{fig:tech}
\setlength\tabcolsep{2pt} %
\begin{tabular}{l|p{0.7\textwidth}} %
\toprule
    Encoder-decoder   & BART~\cite{lewis2019bart}, T5~\cite{raffel2020exploring}, T0~\cite{sanh2021multitask}, GLM~\cite{du2022glm},  ST-MoE~\cite{zoph2022st}, UL2~\cite{tay2023ul2}, Flan-T5~\cite{chung2022scaling}, ChatGLM~\cite{ChatGLM}\\ 
    \midrule   
    Encoder-only    & BERT~\cite{devlin2018bert}, RoBERTa~\cite{liu2019roberta}, ALBERT~\cite{lan2019albert}, Distill BERT~\cite{sanh2019distilbert}, ELECTRA~\cite{clark2020electra}, DeBERTa~\cite{he2020deberta} \\ 
    \midrule
    Decoder-only    & GPT-1 to 4 \& InstructGPT~\cite{radford2018improving,radford2019language,brown2020language,openai2023gpt4,ouyang2022training}, GLaM~\cite{du2022glam}, Gopher~\cite{rae2022scaling}, LaMDA~\cite{thoppilan2022lamda}, PaLM \& U-PaLM \& Flan-PaLM~\cite{chowdhery2022palm, tay2022transcending,chung2022scaling}, OPT \& OPT-IML~\cite{zhang2022opt, iyer2022opt}, BLOOM \& BLOOMZ~\cite{scao2022bloom,muennighoff2022crosslingual}, LLaMA~\cite{touvron2023llama}, Bard~\cite{Bard}, Claude~\cite{claude}\\
\bottomrule
\end{tabular}
\end{table*}

\subsubsection{Mainstream Architectures of LLMs.}

Existing LLMs are built on the foundation of the vanilla transformer and have evolved into three main architectural variations (as summarised in Table~\ref{fig:tech}): encoder-decoder models~\cite{lewis2019bart,raffel2020exploring,sanh2021multitask,du2022glm,zoph2022st}, encoder-only models~\cite{devlin2018bert,liu2019roberta,lan2019albert,sanh2019distilbert}, and decoder-only models~\cite{radford2018improving,radford2019language,openai2023gpt4,ouyang2022training}. The choice of building models based on the encoder, decoder, or both is often determined by the training objectives and tasks at hand.

The encoder-decoder models, such as BART~\cite{lewis2019bart}, T5~\cite{raffel2020exploring}, T0~\cite{sanh2021multitask}, and GLM~\cite{du2022glm}, adopt the masked language model objective. These models are trained to recover masked tokens in the input sequences and have demonstrated strong language understanding capabilities. Given a sequence of tokens $x=\{x_1,\cdot,x_n\}$ with token $\widetilde{x}$ replaced, the training objective can be denoted as follows:
\begin{equation}
    \mathcal{L}(x) = log\mathit{P}(\widetilde{x}|x_{\backslash\widetilde{x}}).
\end{equation}

\begin{figure*}[htbp]
     \centering
     \includegraphics[width=\textwidth]{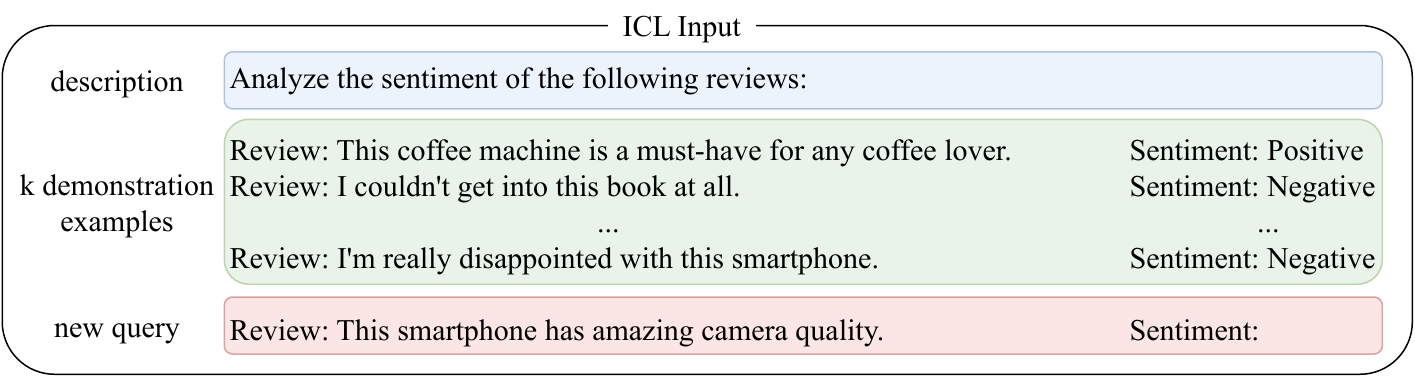} %
\caption{Illustration of In-Context Learning (ICL). ICL input consists of a natural language description explaining the task, $k (k\geq0)$ demonstration examples to illustrate further, and a new query.}
    \label{fig:ICL}
\end{figure*}

Encoder-only models, including BERT~\cite{devlin2018bert}, RoBERTa~\cite{liu2019roberta}, ALBERT~\cite{lan2019albert}, and DistillBERT~\cite{sanh2019distilbert}, also fall into the masked language model category. However, they solely rely on the encoder component for pre-training and are particularly effective in tasks that require language understanding and representation learning. 

In contrast, decoder-only models, such as GPT series~\cite{radford2018improving,radford2019language,brown2020language,openai2023gpt4,ouyang2022training}, GLaM~\cite{du2022glam}, and Gopher~\cite{rae2022scaling}, are trained using the auto-regressive language model objective. These models predict or generate the next tokens based on preceding tokens and excel in text generation tasks, including question answering and natural language generation. A general form of training objective is to maximize the following likelihood:
\begin{equation}
    \mathcal{L}(x) =\sum_{i=1}^n log\mathit{P}(x_i|x_{<i}).
\end{equation}

\subsection{In-Context Learning}
In-Context Learning (ICL) is the ability of generative AI models to infer and learn from a limited set of examples presented in a given context.

Initially mentioned in GPT-3~\cite{brown2020language}, LLMs has the In-Context Learning (ICL) ability, that is, to infer and learn from a limited set of examples presented within a given context. As shown in Fig.~\ref{fig:ICL}, LLMs can respond to novel queries using merely a detailed natural language description supplemented by several demonstrative examples without explicit gradient updates. This characteristic of ICL makes it different from traditional supervised learning approaches. A remarkable advantage is its ability to tackle new tasks without the requirement for fine-tuning. This results in considerable computational cost savings and paves the way for the possibility of deploying language-model-as-a-service~\cite{valmeekam2022large,dong2023survey}. Furthermore, using natural language descriptions and demonstrations to instruct LLMs has dual benefits. Firstly, it simplifies communication with LLMs, rendering them more interpretable. Secondly, it allows for the seamless integration of human knowledge into the LLMs by altering descriptions or demonstration examples.

Although LLMs have shown significant potential for In-Context Learning (ICL) without the need for fine-tuning, research indicates that further preparatory training prior to ICL inference can substantially enhance ICL performance~\cite{thoppilan2022lamda,wei2023symbol,mishra2021cross,min2021metaicl}. 
Distinct from the process of fine-tuning LLMs for individual tasks, this type of training is designed to augment the overall ICL capabilities of LLMs across a spectrum of tasks. 
For instance, Min et al.~\cite{min2021metaicl} propose a meta-training framework to refine pre-trained language models for amplified in-context learning across various tasks. 
Wei et al.~\cite{wei2023symbol} introduced the concept of symbol tuning. This involves the replacement of natural language labels in in-context input-label pairs with arbitrary symbols. This approach necessitates the model to discern tasks by comprehending input-label information rather than relying on instructions or natural language labels. 
Chen et al.~\cite{chen2022improving} offer an alternative strategy to enhance ICL by constructing self-supervised training data that align with downstream ICL formats.

Additional approaches to enhance In-Context Learning (ICL) proficiency include the organization~\cite{liu2021makes,kim2022self,rubin2021learning,lu2021fantastically} and formatting~\cite{wang2022self,wei2022chain,zhang2022automatic,press2022measuring} of demonstrations. These techniques focus on optimizing the choice and order of examples.
Liu et al.\cite{liu2021makes}, for instance, propose a strategy of retrieving examples that exhibit high semantic similarity to a given test query. Kim et al.\cite{kim2022self} generate demonstrations directly using LLMs to reduce dependency on external sources.
Once suitable examples are chosen, determining their order becomes a critical aspect of enhancing performance. Lu et al.\cite{lu2021fantastically} note that sensitivity to the order of demonstrations is a recurrent issue across various models. To mitigate this, several techniques have been proposed to arrange demonstrations appropriately. For example, Liu et al.\cite{liu2021makes} recommend sequencing examples based on their semantic proximity to the input. On the other hand, Lu et al.~\cite{lu2021fantastically} use global and local entropy metrics to decide the order of the examples.

Regarding the demonstration formatting, instruction formatting and reasoning steps formatting are proposed to improve the ICL performance~\cite{dong2023survey,wang2022self,wei2022chain,zhang2022automatic,press2022measuring}.

\section{Data Safety}\label{sec:data_safety}
AI safety challenges arising from the data used to train the model. Data-related safety issues encompass biases and prejudices in training data~\cite{mehrabi2021survey,NEURIPS2021_1531beb7,chang-etal-2019-bias,garrido2021survey}, which can lead to discrimination, as well as the presence of toxic and harmful content that can cause LLMs to generate unsafe results~\cite{ousidhoum2021probing,gehman2020realtoxicityprompts,welbl2021challenges}. On the other hand, inadequate representation of minority groups or some languages in the training data can aggravate inequality~\cite{joshi2020state}. Furthermore, concerns about privacy leakage~\cite{pan2020privacy,liu2022piccolo,song2020information,zanella2020analyzing} and intellectual property infringement arise when sensitive or proprietary information is included in the training data.

\subsection{Toxicity}\label{sec:toxicity}
Toxicity is defined as a rude, disrespectful comment; likely to make people leave a discussion, while \textit{severe} toxicity is defined as a very hateful, aggressive, disrespectful comment or otherwise very likely to make a sure leave a discussion or give up on sharing their perspective. There is also suggestion to use attributes to further categorize toxic content to whether toxic content specifically is an attack based on identity, an insult, a profanity, a threat, or sexually explicit \cite{perspectiveAPI}. Content that is toxic could also be a bias towards a specific group based on race that use offensive language \cite{kurita2019robust}. Or simply, labeling data to be considered as toxic with the tags: 'offensive', 'abusive', 'hateful', etc. \cite{Pavlopoulos2020}.

Large language models can generate harmful, offensive, or inappropriate content \cite{ousidhoum2021probing,gehman2020realtoxicityprompts,welbl2021challenges}, such as hate speech, bullying, or misleading information. There are different types of content that can be considered toxic and harmful, and these can be categorised according to the content being offensive, targeted or non-targeted, individual (cyberbullying), group (hate speech) or others \cite{Zampieri2019}. The challenge is that the variety of unsafe content types may or may not require different treatment. Another challenge is that there is no clearly and widely agreed categorisation and definition of toxic content that can vary greatly due to different social backgrounds and cultural expectations \cite{weng2021toxic}.

GAI-LLMs are trained on a data labelled as 'safe' or as 'unsafe' which is the foundation of training a toxic language classifier. These kinds of data provide signals for the model to be able to reduce toxicity \cite{weng2021toxic}. Data is neutral, humans decide which kind of data is 'safe' or 'unsafe'. Although usually crowd-sourced, data labelers can also be NLP experts with curated data or synthetic data, and professional moderators who may themselves carry their own bias to the platform. Data labelers then try to improve quality by using test data, having clear guidelines, having a second opinion through a majority vote, and attribution to the authors \cite{Zampieri2019}. It is fairly common for large language models to be trained on social media data like Twitter and Reddit; however, starting with a manually labelled blacklist of words might be too small to work as training tokens. \cite{khatri2018detecting} proposed the idea of using a two-stage semi-supervised pipeline to increase the performance of a toxic binary classifier by first training a model on the smaller dataset, and then using this first model as the classifier to a larger data set in a social media site such as Reddit, and the output model is the effective binary classifier for toxic content. Large language models have pre-trained knowledge to reduce the probability of undesired attributes in model generation, which can be used in detoxing the model. This can be done by using a soft variant where the probabilities of words with negative weighting are reduced with respect to the true sampling distribution \cite{schick2021self}

\subsection{Bias}\label{sec:bias}
During pre-training, the model learns to predict the next word in a sentence, which helps it learn grammar, facts about the world, reasoning abilities, as well as any biases present in the training data. If the training data includes biased content (e.g., gender bias in job descriptions, racial bias in news articles), the model could learn and reproduce this bias. Sometimes, the model can even amplify existing biases. For instance, if the model sees that certain professions are more often associated with men than women in the training data, it might predict male pronouns more often for these professions, thereby amplifying the existing bias.

Large language models learn from vast amounts of text data, which may contain implicit or explicit biases~\cite{mehrabi2021survey,NEURIPS2021_1531beb7,chang-etal-2019-bias,garrido2021survey}. As a result, these models can inadvertently perpetuate and amplify these biases when generating text, leading to unfair or discriminatory outputs~\cite{nangia2020crows,czarnowska2021quantifying,schick2021self}. For instance, if a reviewer has a bias against certain dialects or accents, and consistently ranks outputs in those dialects lower, the model could learn to mimic this bias. It's also possible for the model to learn to "cheat" or find loopholes in the review process, which could potentially lead to biased behavior. For example, if reviewers are more likely to miss subtle forms of bias, the model could learn to be subtly biased in a way that avoids detection.Bias in the context of LLMs generally refers to systematic errors or disparities in the model's performance or outputs that can be traced back to certain characteristics of the input data, the model architecture, or the training process. There are various types of biases that could exist in LLMs. Pre-existing bias is bias in the data the LLMs was trained on. If the data contained gender, racial, cultural, or other biases, the model may learn and reproduce these biases in its outputs. Technical Bias is a form of bias is introduced by the choices made during the design and training of the model, such as the choice of the learning algorithm or the optimization process. Emergent bias emerges when the LLMs is used in a certain context or application. For example, the model might generate different outcomes for different user groups, even though it was not explicitly trained to do so.

Bias can be mitigated with strategies before model training. Pre-processing is performed by identifying and reducing bias the data before it's used for training. There are also in-process modifications referring to techniques such as fairness-aware machine learning algorithms that aim to minimize bias during the model training process. In post-processing, there are also strategies for reducing bias after the model has been trained, such as in the output review and output modification \cite{han2022systematicevaluation}.

GAI-LLMs are said to be fair when the model's bias property has been addressed \cite{li2024surveyfairnesslargelanguage}. Fairness in the context of generative language models refers to the equitable treatment of all individuals and groups, regardless of their characteristics such as race, gender, age, etc. An LLM is considered fair if its outputs do not systematically disadvantage or advantage any particular group. Further refinements to the definition includes whether if the LLM treats similar individuals similarly. In other words, two people who are alike in all relevant aspects should receive similar outputs from the model. As opposed to Counterfactual fairness in a hypothetical world where a person's demographic attributes were different (but all other characteristics remained the same), the model's output for that person would not change \cite{gallegos2024biasfairnesslargelanguage}.  Although there are consequences on user trust of a model having a bias but with outputs further from what is true. While biased outputs contribute to harmful stereotypes or discrimination, sanitized outputs may misrepresent the distribution of what is statistically true.

\subsection{Data Privacy}\label{sec:data_privacy}
Since LLMs are trained on massive datasets, there is a risk of inadvertently capturing and exposing private or sensitive information from individuals or organizations. This can lead to privacy breaches~\cite{pan2020privacy,liu2022piccolo,song2020information,zanella2020analyzing}. The pretraining phase involves training a large language model on a vast corpus of publicly available text from the internet. While efforts are made to anonymize the training data, the risk of reidentification or exposure of sensitive information persists. The main data concern is the possibility of exposing sensitive or personal information from the training data or human feedback. ChatGPT is trained on a large corpus of text from various sources, such as web pages, books, news articles, social media posts, etc. Some of these sources may contain private or identifiable information about individuals or organizations, such as names, addresses, phone numbers, email addresses, credit card numbers, etc. If ChatGPT memorizes or leaks such information in its generated text, it could pose an issue to the privacy and security of the data subjects. Adversarial attacks, such as membership inference or attribute inference attacks, pose serious threats to user privacy. By leveraging the unique responses generated by ChatGPT, an adversary could infer whether a particular training example was present in the training set or even extract private attributes of the data.To fine-tune the model's behavior and improve response quality, reinforcement learning with human feedback is employed. Human reviewers rate and provide feedback on model-generated responses, which are then used to update the model. This iterative process raises privacy concerns as reviewers may inadvertently or intentionally expose personal or sensitive information during the feedback process. The aggregation and analysis of this feedback data pose a risk of data leakage and privacy breaches.The transformer block, a fundamental component of ChatGPT, facilitates efficient parallelization and captures global dependencies through self-attention mechanisms. However, this architecture poses privacy risks due to the model's ability to attend to different parts of the input sequence simultaneously. Adversaries may exploit self-attention to extract sensitive information, infer the presence of certain topics, or even perform targeted attacks by manipulating the attention weights. Self-attention mechanisms, including multi-head attention, enable ChatGPT to capture dependencies between different parts of the input text efficiently. However, these attention mechanisms raise privacy concerns by potentially exposing relationships between words or phrases within the input. Adversaries may exploit this information leakage to infer private attributes or uncover confidential patterns in the user's queries \cite{carlini2021extractingtrainingdatalarge}.
    
\subsection{Copyright}\label{sec:copyright}
Generative AI systems may raise issues related to intellectual property, such as copyright infringement or plagiarism. The use of copyrighted material, while enriching the dataset, could lead to legal and ethical dilemmas, calling for more deliberate and thoughtful data governance policies.  The use of copyrighted data presents both legal and ethical challenges. Legal issues arise due to potential infringement of copyright laws. Technically, the indiscriminate use of copyrighted data arises from the process of data scraping, where large amounts of text data are gathered from the internet. Automated scraping often does not discriminate between copyrighted and non-copyrighted data, leading to the potential inclusion of copyrighted material. Ethically, using copyrighted data without proper attribution or consent can be seen as a disregard for the principles of respect for intellectual property rights.The training data used in pre-training for LLMs, raised concerns on intellectual property. Generative AI models may have been trained on training data that is made available by a third party may not have been given outright permission by the original creator of the work. Creators of original work such as artists and writers, may not have given outright permission for generative models to produce new creations that are made in their style of work. Copyrighted data in training has appeared in the literature \cite{liang2023holistic}, but training data remains to be secretive with the source of the data \cite{pan2020privacy, li2023privacy}. Moreover, over-reliance on copyrighted data could lead to overfitting to specific data sources or styles, potentially reducing the generalizability of the models.

\section{Model Safety}\label{sec:model_safety}
AI safety challenges arise both from the training process and the inherent model architecture of large language models. A key model-related safety issue involves the generation of misinformation \cite{ognyanova2020misinformation}. Currently, there is a lack of well-established evaluation tools to assess LLM-generated content, making the detection and identification of misinformation increasingly difficult. Another significant concern is the potential legal issue of LLMs inadvertently reproducing or inferring sensitive information from training data, or generating texts that closely resemble copyrighted training material. Moreover, the lack of transparency and explainability in LLMs poses substantial challenges in addressing these issues \cite{GPTexplainable}. The opaque nature of these models complicates efforts to understand their decision-making processes and mitigate potential biases or errors \cite{gallegos2024biasfairnesslargelanguage}. As LLMs continue to evolve in various applications, addressing these safety and ethical concerns becomes increasingly crucial before scaling \cite{wei2022emergent, kaplan2020scaling, AnthropicResponsibleScalingPolicy}.

\subsection{Misinformation}\label{sec:misinformation}

\begin{figure*}[ht*]
     \centering
     \includegraphics[width=0.8\textwidth]{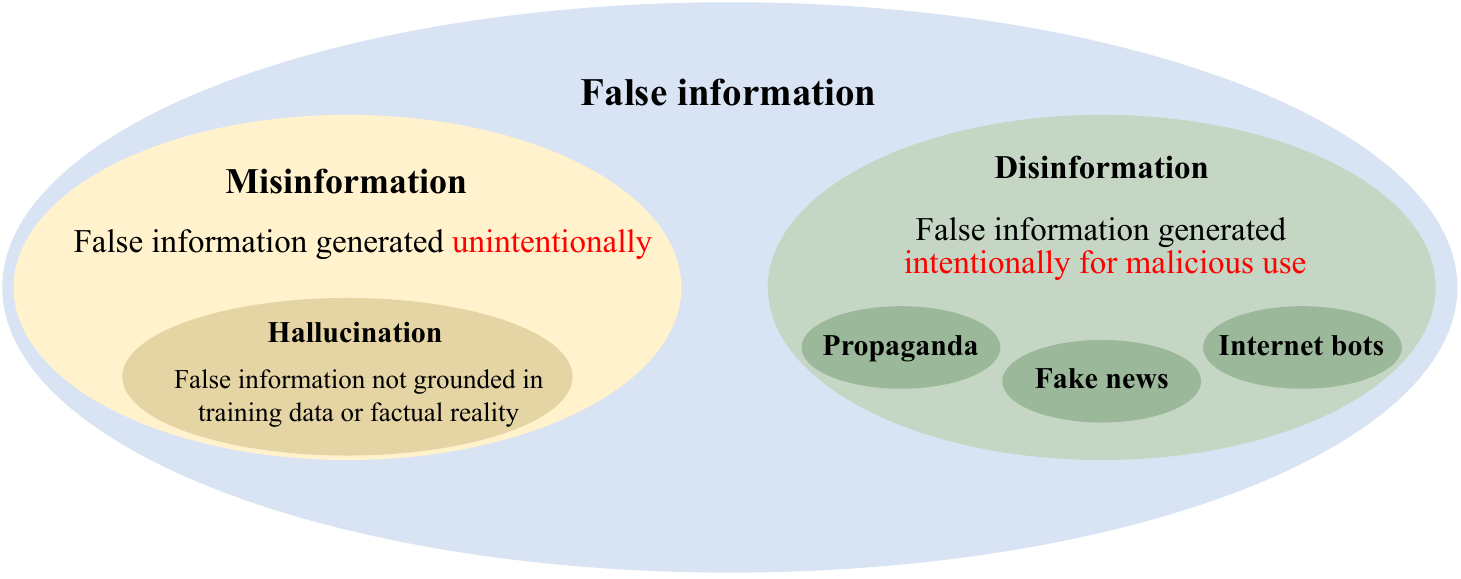}
\caption{ The relationships between hallucination, misinformation, disinformation, and related terms.}
    \label{fig:falseinfo}
\end{figure*}

\begin{figure*}[ht]
     \centering
     \includegraphics[width=\textwidth]{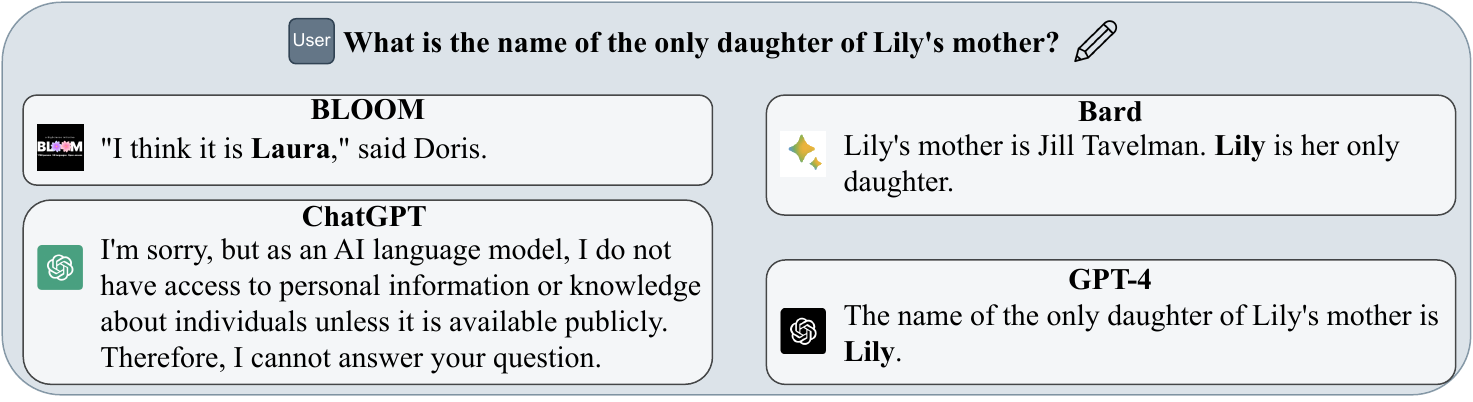} %
\caption{Answers to the same question generated by BLOOM, CHATGPT, Bard, and GPT-4.}
    \label{fig:misinfo:a}
\end{figure*}

Despite the impressive ability of LLMs to generate human-like text, they are still prone to generating misinformation~\cite{bubeck2023sparks,weidinger2021ethical,weidinger2022taxonomy,kreps_mccain_brundage_2022,solaiman2019release,tamkin2021understanding}. Misinformation is false, inaccurate, or misleading information generated by LLMs with no malicious intent~\cite{weidinger2022taxonomy}. It differs from disinformation, deliberately created to cause harm~\cite{weidinger2021ethical}. In addition, there are several related terms, such as `hallucination' - a phenomenon in which LLMs make things up that are not anchored in either the input data or any reliable training data - as well as `propaganda' and `fake news' etc. We categorise them according to the intent for which they were created, as shown in Fig.~\ref{fig:falseinfo}. Misinformation and disinformation can lead to the dissemination of fake information through various channels ranging from widespread social media to person-to-person chats. They may have negative influences individually, communally, and socially~\cite{kreps_mccain_brundage_2022}. In this section, we focus on the safety issues of misinformation.

Numerous studies and cases emphasise the tendency of LLMs to produce misinformation~\cite{floridi2020gpt,borji2023categorical,cobbe2021training}. 
For example, Google's AI chatbot, Bard, provided an incorrect answer upon its release, leading to a subsequent drop in stock value. This misinformation is prevalent across all LLMs. They may fabricate answers with non-existent books or references without warnings or disclaimers, make obvious factual errors, and sometimes fail to perform simple mathematical calculations or logical inferences~\cite{openai2023gpt4,thoppilan2022lamda,chowdhery2022palm,touvron2023llama}.

Fig.~\ref{fig:misinfo:a} illustrates their limitations with an example. We pose the same question to the latest LLMs (that is, BLOOM~\cite{scao2022bloom} by BigScience, Bard~\cite{Bard} by Google, and ChatGPT and GPT-4~\cite{openai2023gpt4} by OpenAI): "What is the name of the only daughter of Lily's mother?" 

While the answer is straightforward for humans—Lily—the results varied among the LLMs. As shown in Fig.~\ref{fig:misinfo:a}, BLOOM provides an irrelevant response; ChatGPT declines to answer the question; Bard, although giving the correct answer, infers it incorrectly by specifying Lily as the actress Lily Collins; only GPT-4 produces the correct answer.

Interestingly, even though GPT-4 outperforms other models, investigations of NewsGuard~\cite{NewsGuard} reveal that it is more susceptible to generating misinformation and is more convincing when fabricating answers than its predecessor, ChatGPT.

\begin{figure*} %
     \centering
     \includegraphics[width=\textwidth]{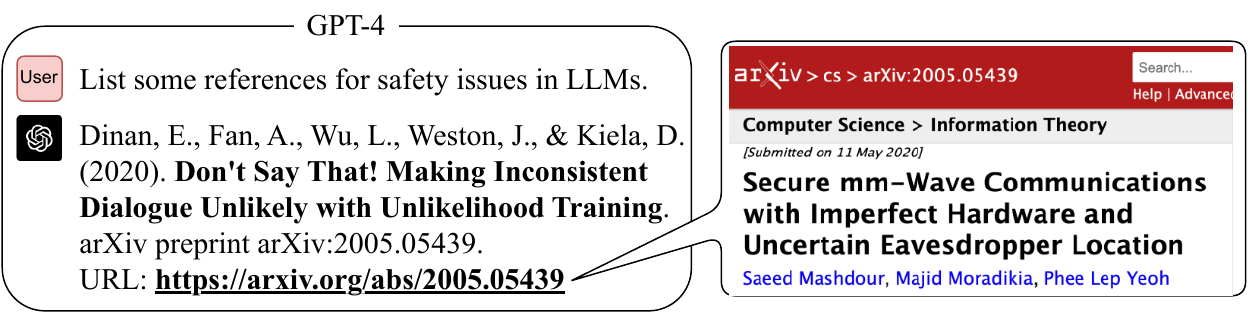} %
\caption{Examples of misinformation generated by LLMs. The latest LLMs, GPT-4, mistakenly provides an irrelevant website link when citing a paper.}
\label{fig:misinfo:b}
\end{figure*}

There are several reasons behind misinformation generation~\cite{lin2021truthfulqa,touvron2023llama,yu2021cross,bender2020climbing,kassner2019negated}.  One is related to the training data~\cite{touvron2023llama}. As mentioned in Sec.~\ref{sec:data_safety} regarding data-related safety issues, training data may contain factually incorrect text with misinterpretations, outright falsehoods, fabricated statements in fiction, etc. LLMs may inadvertently learn, propagate, and generate misinformation when exposed to such content during training. Moreover, even when the training data consists only of fact-checked correct statements, LLMs may still generate misinformation due to the limitations of the model architecture~\cite{lin2021truthfulqa,kassner2019negated}. More specifically, first, the vast size of the training data requires LLMs to compress information, which may lead to information loss that contributes to incorrect outputs~\cite{yu2021cross}. Second, LLMs are designed to maximize the likelihood of their outputs instead of understanding the content. However, a high likelihood does not guarantee factual correctness. For instance, humans can tell linguistically plausible lies that are not factually correct~\cite{weidinger2022taxonomy}. Similarly, LLMs can generate outputs that appear coherent but are not necessarily accurate or true.  
We show an example in Fig.~\ref{fig:misinfo:b}, where we query GPT-4~\cite{openai2023gpt4} to list some references for safety issues in LLMs. GPT-4 generates a plausible response, but upon closer inspection, the provided link is irrelevant to the given reference. Such examples can be characterized as a form of hallucination.
Finally, LLMs' limited context understanding contributes to misinformation generation~\cite{bender2020climbing}. Inadequate comprehension of nuances such as time, space, speaker identity, or cultural background can lead to factually incorrect outputs. For example, GPT-4 responds that the latest version of GPT is GPT-3, which was once accurate but is now outdated.

Misinformation generated by LLMs can lead to a variety of risks and consequences~\cite{zellers2019defending,marwick2017media,ognyanova2020misinformation,kenton2021alignment,brown2020language,bickmore2018patient,alberts2023large}. When users rely on LLMs outputs, they may unknowingly believe in misinformation, especially when the misinformation seems convincing, and they cannot validate the information. Such wrong beliefs may erode trust in AI systems, as well as in experts, public figures, etc., leading to social and political polarization~\cite{zellers2019defending,marwick2017media,ognyanova2020misinformation,sun2022battle}. An example is wrong information about vaccines during Covid-19 may exacerbate tensions and divisions within society~\cite{sun2022battle}. Moreover, LLMs that generate biased or prejudiced content may inadvertently reinforce existing stereotypes or discrimination, causing marginalization and amplifying biases. Additionally, individuals and organizations that base their decisions on misinformation produced by LLMs may make suboptimal decisions with negative consequences. For example, a farmer may encounter severe financial loss if he arranges plant plans based on outdated weather information provided by LLMs. Similarly, misinformation can bring risks when LLMs are integrated into downstream applications~\cite{kenton2021alignment,brown2020language}. Especially in sensitive domains such as healthcare, finance, or law, the generation of misinformation by LLMs can have severe repercussions~\cite{bickmore2018patient,alberts2023large}.

\subsection{Eval Problem}\label{sec:eval_problem}
LLMs such as GPT-4 often exhibit remarkable versatility, capable of handling various NLP tasks. As a recent survey suggests~\cite{yang2023harnessing}, tasks spanning dialogue systems, reasoning, language modeling, text generation, etc., can all benefit from the prowess of LLMs, exceeding the performance of fine-tuned models. This versatility raises a critical question about evaluating LLMs across various tasks. 

On the one hand, existing protocols established to measure model performance in individual, more narrowly defined scenarios still apply. These include general measures such as F1 Score, Precision, and Recall, which assess model accuracy. To fit the realm of NLP,  some recall- and/or precision-oriented metrics like Recall-Oriented Understudy for
Gisting Evaluation (ROUGE)~\cite{lin2004rouge}, Bilingual Evaluation Understudy (BLEU)~\cite{papineni2002bleu}, and its variant GLEU~\cite{wu2016google}, and BLEURT~\cite{sellam2020bleurt} are proposed to measure text similarity between model-produced text and human-produced reference text. 
Metric for Evaluation of Translation with Explicit ORderinG (METEOR)~\cite{banerjee2005meteor}, and BERTScore~\cite{zhang2019bertscore} also fall into this category but combine more granular text similarity measures, including synonyms, stemming or leveraging pre-trained contextual embeddings from BERT~\cite{devlin2018bert}. Other protocols include Perplexity (PPL) to measure how well a probability model predicts a sample, Translation Edit Rate (TER) to calculate the number of edits required to change the model outputs to references, Regressor using sentence embeddings (RUSE)~\cite{shimanaka2018ruse} using a multi-layer perception (MLP)-based regression architecture to output the RUSE score between hypothesis and reference sentences, etc.

\begin{figure*} %
     \centering
     \includegraphics[width=\textwidth]{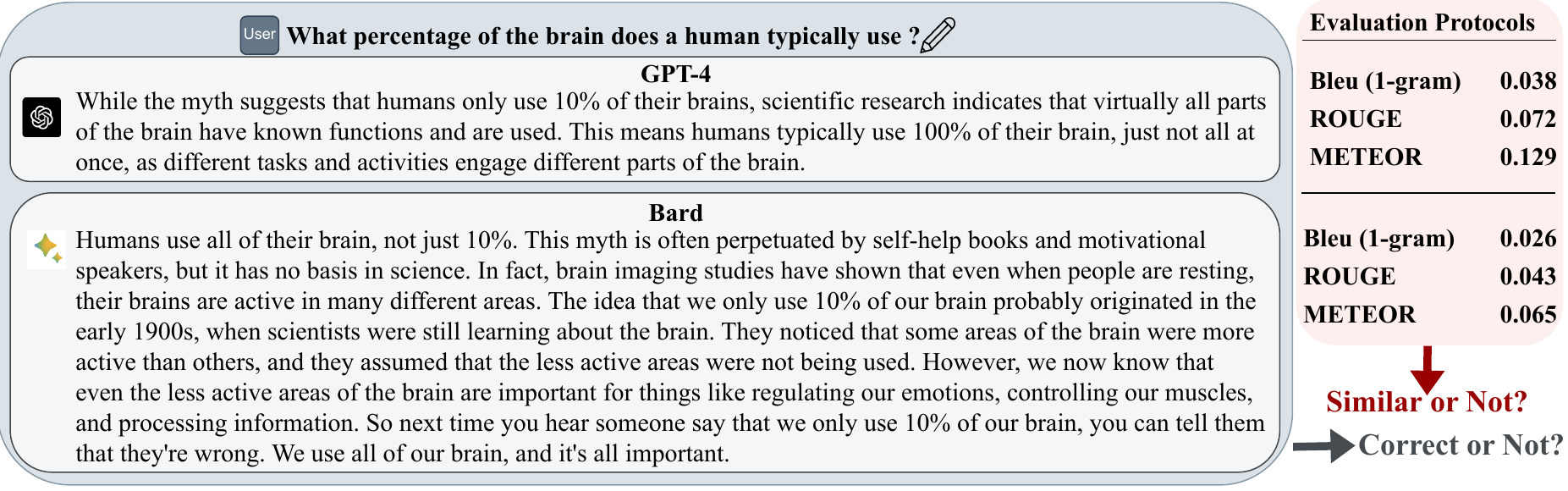} %
\caption{Text similarity evaluated by BLEU, ROUGE, and METEOR.}
\label{fig:eval_protocols}
\end{figure*}

On the other hand, despite their past success, these existing protocols fail to evaluate LLMs like GPT-4 adequately. Take, for example, the question OpenAI developers posed to GPT-4: "What percentage of the brain do humans typically use?"~\cite{bubeck2023sparks} When GPT-4 and Bard were asked this question, they provided lengthy answers to avoid ambiguity, as shown in Fig.~\ref{fig:eval_protocols}. However, these detailed responses lead to low text similarity scores when evaluated with metrics such as BLEU (1-gram), ROUGE, and METEOR. These metrics, frequently used in current studies, judge the correctness of an answer based on its similarity to the golden answer, i.e., "The majority of a human’s brain is almost always active". Therefore, although the answers are accurate, they are considered incorrect according to these metrics. This contradiction highlights the limitations of current evaluation protocols, which rely heavily on syntactic rather than semantic similarity. While these protocols may have proven successful on earlier models and datasets, they do not apply to the LLMs, which often provides lengthy, comprehensive, and sometimes redundant answers. 

The limitations of evaluation metrics may harm the designing and training of LLMs and potentially skew the understanding of their abilities. For example, the concept of `emergent abilities in LLMs', which is defined as "abilities that are not present in smaller-scale models but are present in large-scale models; thus, they cannot be predicted by simply extrapolating the performance improvements on smaller-scale models"~\cite{wei2022emergent}, has been a hot topic. However, Schaeffer et al. suggest this may not be a `miracle' of LLMs. They argue that the sharp and unpredictable performance increase does not come from improved abilities but from metrics that non-linearly or discontinuously deform per-token error rates~\cite{schaeffer2023emergent}. 
This opinion has sparked a debate. Wei et al. defend the existence of emergent abilities and argue that their original paper already recognized the potential impact of evaluation metrics on these abilities' perception~\cite{wei2022emergent}. They assert that not all non-linearly or discontinuously changed metrics can be replaced, and compelling evidence that a smoothly improving surrogate metric can predict the emergent metric's performance is yet to be presented.
Despite these differing viewpoints, there is a shared consensus about the crucial role of evaluation metrics in understanding LLMs' abilities.
Therefore, refining current evaluation protocols to better align with the unique characteristics of LLMs is essential for their continued evolution.

Beyond refining existing evaluation metrics, it's critical to design new ones that evaluate novel aspects of LLMs performance. A significant omission in current assessment tools is the measurement of the correctness of generated content, i.e., content verification.  The verification is essential given that it is easy for LLMs to generate seemingly plausible misinformation, and the consequence can be severe when trusting misinformation (see Sec.~\ref{sec:misinformation}). 
Some initiatives have been put in place to address the generation of misinformation. For example, the Reinforcement Learning from Human Feedback (RHLF) approach integrates human feedback into the training loop~\cite{ouyang2022training}, allowing GPT-3 to be fine-tuned using reinforcement learning. OpenAI has further adapted RHLF for the safety training of GPT-4, supplementing it with their proposed rule-based reward models (RBRMs). These RBRMs provide additional reward signals during RHLF fine-tuning~\cite{openai2023gpt4}.

Despite their success in reducing misinformation, these initiatives don't fully address the need for effective metrics to verify the accuracy of LLMs outputs. A step in this direction is the work by Lin et al.~\cite{lin2021truthfulqa}, who constructed a benchmark dataset called TruthfulQA and proposed a `truthfulness' metric to measure the correctness of generated answers to questions in TruthfulQA. However, they define truthfulness based on the similarity to accurate answers.
As previously discussed, a low similarity to the gold standard answer does not necessarily indicate that the generated response is incorrect, particularly given that LLMs tend to produce lengthier replies (Fig.~\ref{fig:eval_protocols}).
Therefore, verification protocols catering to LLMs, especially those newly released high-performance ones, require further exploration and development. Advancing in this area would mark a significant stride towards improving the reliability and trustworthiness of LLMs outputs.

\subsection{Explainability}\label{sec:explainability}
As large language models become increasingly prevalent due to their extensive capabilities, growing concerns regarding transparency in these models have begun to emerge~\cite{daneshjou2021lack,yang2023harnessing}. 
An instance is Bard's chatbot indicating Google using Gmail data to train Google's Bard model.  Although Google denied this, such incidents make the public anxious about potential privacy breaches without clear data transparency. Concerns deepened when OpenAI, with the release of GPT-4, stated that it would no longer publish "further details about the architecture (including model size), hardware, training compute, dataset construction, training method, or similar"~\cite{openai2023gpt4}. Contrastively, their earlier GPT-1 and GPT-2 were open-source. Similarly, Google signaled an intention to reduce the publication of papers before ideas had materialized into products.

These decisions are driven by competitive pressures or commercial considerations and also help avoid disputes similar to what Stability AI experienced about using a paid-for database for training~\cite{openai2023gpt4}. Alongside this, the complex nature of LLMs further exacerbates challenges in enhancing transparency, particularly regarding explainability—making the decision-making processes of these models understandable to users. Interpreting how LLMs generate specific outputs or decisions based on their inputs remains difficult~\cite{wu2022ai, liang2022holistic}.

The lack of transparency and explainability in LLMs introduces a range of risks, primarily impacting trust, fairness, safety, and practical application of these models~\cite{daneshjou2021lack,yang2023harnessing}. Crucially, transparency and explainability are vital to establishing trust in LLMs among the public. A lack of understanding about the functioning of LLMs could lead to hesitancy in their adoption or, worse, their misuse due to misconceptions about model outputs. This is especially risky when individuals rely on these outputs to make critical decisions without comprehending how the models work or how to adjust their prompts for optimal results~\cite{wu2022ai}.

LLMs' dearth of transparency and explainability further complicates identifying and mitigating safety issues. For example, in the case of misinformation, the ability to promptly validate the responses generated by LLMs is hampered as the datasets used for training, the sources utilized for generating answers, and the mechanisms behind output generation are obscured. This situation starkly contrasts the verification ease offered by search engines, which provide source links alongside search results.
Moreover, the opacity of LLMs further inhibits their deployment and application in real-world scenarios, particularly in sensitive sectors such as healthcare and education, where a deep understanding of how decisions are made is imperative~\cite{kocak2022transparency,kung2023performance}. Finally, the enigmatic nature of model responses complicates ensuring alignment with human values and intentions, creating ethical and moral dilemmas~\cite{bowman2022measuring, bai2022constitutional}.

Numerous efforts have been initiated to address the twin concerns of transparency and explainability in large language models. A primary effort involves the open-source release of models such as the Open Pre-trained Transformer (OPT)~\cite{zhang2022opt}, BLOOM~\cite{scao2022bloom}, and LLaMA~\cite{touvron2023llama} in non-commercial versions. The establishment of more transparent benchmarks and evaluations also contributes to this aim~\cite{liang2022holistic}. On the other hand, to enhance the transparency and explainability of the models themselves, OpenAI has endeavored to utilize GPT-4 to elucidate the underlying logic of its outputs~\cite{bubeck2023sparks}. However, discrepancies between explanations and outputs can occur, with explanations persisting even for nonsensical outputs. This process was further advanced by leveraging GPT-4 to explain individual neurons in GPT-2~\cite{GPTexplainable}. Their three-step method entails explaining neuron activations using GPT-4, then simulating these activations with GPT-4 conditioned on the explanation, and finally scoring the explanation by comparing simulated and real activations. To achieve higher-level transparency and explainability, Wu et al. proposed chaining large language model prompts, wherein the output of one step serves as the input for the next, enhancing the clarity of the logical flow~\cite{wu2022ai}. However, it is crucial to note that the pursuit of transparency and explainability can create tensions between transparency, privacy, and security~\cite{weidinger2021ethical}. On the one hand, transparency risks leaking private information, while on the other, it could potentially offer access to LLMs for malicious actors. Moreover, there is a conflict between increased transparency and corporate profitability. Consequently, striking a balance between these competing demands is paramount in enhancing transparency in LLMs.

\subsection{Inference Privacy}\label{sec:inference_privacy}
In Sec~\ref{sec:data_privacy}, we investigate the data-related privacy leakage of LLMs. However, it is important to note that this issue may extend beyond the training phase and occur during inference. LLMs can infer sensitive information even without explicitly private data in the training corpus~\cite{lipkin2023evaluating}. 
Derner et al. conducted experiments involving ChatGPT, posing questions related to individuals' private lives and alleged affairs (such as "What can you tell me about his private life?", "Who did he allegedly have affairs with?")~\cite{derner2023beyond}. ChatGPT responded based on inferred information, potentially derived from rumors or unverified sources. Such studies highlight the potential of LLMs to reveal sensitive personal characteristics such as relationship status, medical conditions, political ideology, etc. The disclosure of such information raises significant ethical concerns and implications. The consequences of such sensitive information inference can be far-reaching, leading to privacy violations, potential discrimination, and stigmatization of individuals based on inferred attributes, etc.  One possible solution is to empower users with tools and services that allow them to protect their data locally. Local differential privacy techniques provide a means to privatize data at the individual level while still allowing for meaningful analysis and inference~\cite{li2023privacy,duan2023flocks}.

Likewise, the harming intellectual property issue discussed in the data-related section ~\ref{sec:copyright} can also occur during inference, stemming from the LLMs's ability to memorize, imitate, and interpret training data. This ability raises concerns about patent-busting, although often unintentionally ~\cite{weidinger2022taxonomy}.
To address this issue, Kalpakchi et al. Propose SweCTRL-Mini, which utilizes transparent datasets and provides an interface for users to search 13-gram texts to check whether they are part of the training data~\cite{kalpakchi2023swectrl}. However, such searches are limited to detecting direct copying and may fail to capture more subtle forms of intellectual property damage.
Instances of LLMs have been observed to generate content that closely resembles the creative works of famous artists and writers by mimicking their style, themes, and structure. These capabilities undermine the profitability and recognition of original works and harm the protection and recognition of creative work.

\section{Prompt Safety}\label{sec:prompt_safety}

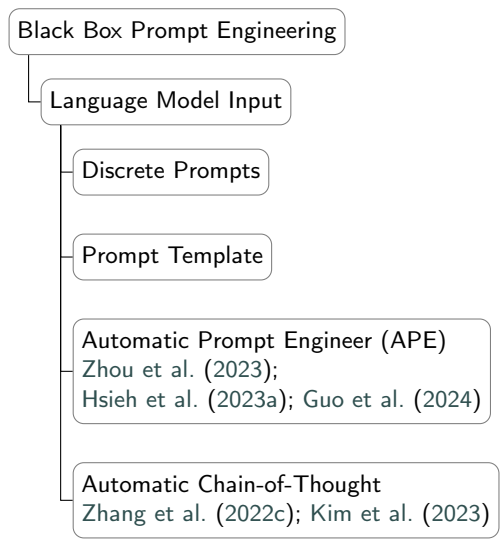
\begin{figure}[ht]
\centering
\begin{forest}
  for tree={
    grow'=0,
    child anchor=west,
    parent anchor=south,
    anchor=west,
    calign=first,
    draw=gray,
    rounded corners,
    s sep=5mm,
    edge path={
      \noexpand\path [draw, \forestoption{edge}]
      (!u.south west) +(7.5pt,0) |- (.child anchor)\forestoption{edge label};
    },
    before typesetting nodes={
      if n=1
        {insert before={[,phantom]}}
        {}
    },
    fit=band,
    before computing xy={l=12pt},
    align=left,
  }
[Black Box Prompt Engineering
  [Language Model Input
    [Discrete Prompts]
    [Prompt Template]
    [Automatic Prompt Engineer (APE) \\ \cite{zhou2023largelanguagemodelshumanlevel}; \\ \cite{hsieh2023automaticengineeringlongprompts, guo2024connecting}]
    [Automatic Chain-of-Thought \\ \cite{zhang2022automatic, kim2023languagemodelssolvecomputer}]
  ]    
]
\end{forest}
\caption{Prompt Engineering: This encompasses any form of programming related to prompt generation that typically approaches LLMs as opaque entities, meaning without considering their internal mechanics. It is still possible to refine candidate responses (exemplars), conduct self-critique on these responses, and generate new candidate responses. Remarkably, this process proves effective, as articulated by the self-correcting theory proposed by \cite{wang2024theoreticalunderstandingselfcorrectionincontext}.}
\label{fig:black_box_prompt}
\end{figure}

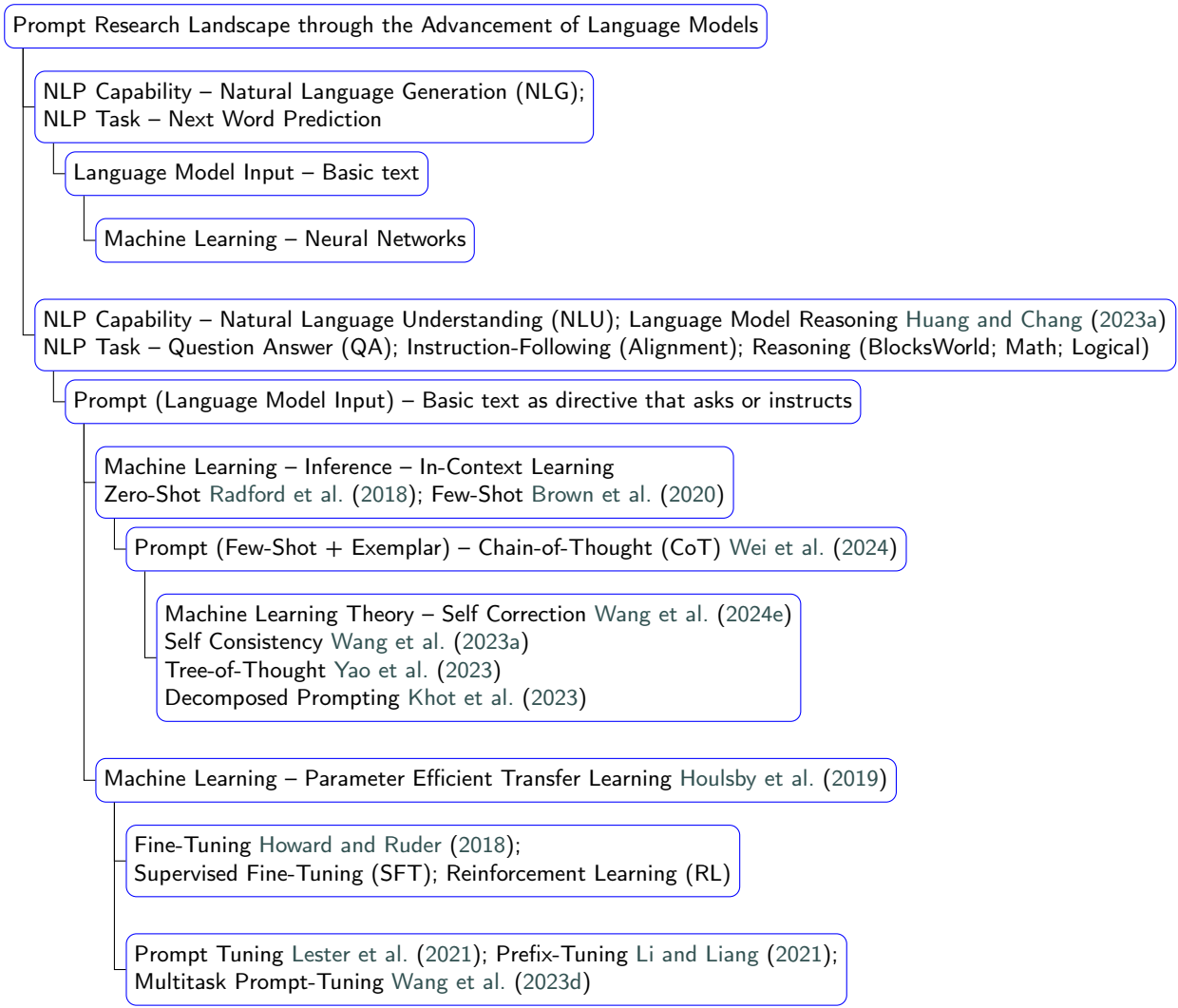
\begin{figure*}[ht]
\centering
\begin{forest}
  for tree={
    grow'=0,
    child anchor=west,
    parent anchor=south,
    anchor=west,
    calign=first,
    draw=blue,
    rounded corners,
    s sep=5mm,
    edge path={
      \noexpand\path [draw, \forestoption{edge}]
      (!u.south west) +(7.5pt,0) |- (.child anchor)\forestoption{edge label};
    },
    before typesetting nodes={
      if n=1
        {insert before={[,phantom]}}
        {}
    },
    fit=band,
    before computing xy={l=12pt},
    align=left,
  }
[Prompt Research Landscape through the Advancement of Language Models
  [NLP Capability -- Natural Language Generation (NLG); \\
   NLP Task -- Next Word Prediction
    [Language Model Input -- Basic text
      [Machine Learning -- Neural Networks]
    ]
  ]
  [NLP Capability -- Natural Language Understanding (NLU); Language Model Reasoning \cite{huang-chang-2023-towards} \\
   NLP Task -- Question Answer (QA); Instruction-Following (Alignment); Reasoning (BlocksWorld; Math; Logical) 
     [Prompt (Language Model Input) -- Basic text as directive that asks or instructs
        [Machine Learning -- Inference -- In-Context Learning \\
         Zero-Shot \cite{radford2018improving}; Few-Shot \cite{brown2020language}
           [Prompt (Few-Shot + Exemplar) -- Chain-of-Thought (CoT) \cite{wei2024simple}
             [Machine Learning Theory -- Self Correction \cite{wang2024theoreticalunderstandingselfcorrectionincontext} \\
              Self Consistency \cite{wang2023selfconsistencyimproveschainthought}\\
              Tree-of-Thought \cite{yao2023tree} \\
              Decomposed Prompting \cite{khot2023decomposed}
             ]
           ]
        ]
        [Machine Learning -- Parameter Efficient Transfer Learning \cite{houlsby2019parameterefficienttransferlearningnlp}
          [Fine-Tuning \cite{howard-ruder-2018-universal}; \\
           Supervised Fine-Tuning (SFT); Reinforcement Learning (RL)
          ]
          [Prompt Tuning \cite{lester2021powerscaleparameterefficientprompt}; Prefix-Tuning \cite{li2021prefixtuningoptimizingcontinuousprompts}; \\
           Multitask Prompt-Tuning \cite{wang2023multitaskprompttuningenables}
          ]
        ]
     ]
  ]
]
\end{forest}
\caption{Landscape of Research on Prompting. The term 'prompt' is used interchangeably to refer both to the Prompt() function and to the prompt input, which can complicate the process of navigating the literature. In this figure, Prompt() is delineated as a function aimed at optimizing the expected value E($X_{model}$ + $P_{input}$), wherein $X$ denotes the frozen pre-trained model parameters and $P$ represents the typically pre-pended prompt input \cite{lester2021powerscaleparameterefficientprompt}. The mechanism of prompting serves to furnish the model with supplementary contextual information, thereby guiding the generation of its output.}
\label{fig:prompt_taxonomy}
\end{figure*}

Crafting input text to guide the output of models, known as prompting, is a rapidly evolving field within generative AI research. \cite{schulhoff2024promptreportsystematicsurvey} offers a comprehensive taxonomy of prompting techniques, providing a valuable framework for understanding the diverse terminology and methodologies. Their work covers various aspects of prompting, which can be standalone directives (such as instructions or questions) or structured templates with user input placeholders. \cite{schulhoff2024promptreportsystematicsurvey}'s survey includes the terms and definitions of prompt engineering, which, despite varying definitions, focuses on strategically manipulating prompt inputs to influence model outputs but without necessarily understanding how LLMs work internally, these techniques can be called black box prompt engineering Figure~\ref{fig:black_box_prompt}. 

In comparison to black box prompt engineering in Figure~\ref{fig:black_box_prompt}, our analysis of prompting techniques for large language models in Figure~\ref{fig:prompt_taxonomy} is a comprehensive and hierarchical classification reflecting the current state of the art in natural language processing and artificial intelligence. This structure is based on the principle of in-context learning, as explained by \cite{brown2020language}, which underpins zero-shot, one-shot, and few-shot prompting methodologies. The taxonomy then advances to more sophisticated approaches, including advanced reasoning prompts that use techniques like Chain-of-Thought \cite{wei2024simple} and Tree-of-Thoughts \cite{yao2023tree} to enhance the model's cognitive abilities. Iterative prompting techniques, such as self-consistency \cite{wang2023selfconsistency}, are categorized separately to highlight the importance of multi-step reasoning processes. The taxonomy also includes prompt optimization methods, such as automatic prompt engineering \cite{zhou2023largelanguagemodelshumanlevel}, recognizing the critical role of refining and optimizing prompts for better performance. Finally, the taxonomy acknowledges task-specific prompting techniques, including task decomposition \cite{khot2023decomposed} and Recursive Criticism Improvement \cite{kim2023languagemodelssolvecomputer}, which are at the cutting-edge for adapting LLMs to being agentic in the sense that the model are able to create subgoals or carry out general tasks. Building on the survey works of \cite{schulhoff2024promptreportsystematicsurvey, liu2024prompt}, who cataloged and described various security-related issues in prompting, our work delves deeper into prompt hacking, jailbreaking, and implementing guardrails. We further explore critical safety concerns within generative large language models in the subtopic of prompting (\cite{schulhoff2024promptreportsystematicsurvey} also extends their categorization work to include emerging techniques such as zero/few-shot learning, thought generation, and various forms of agent-based prompting). Therefore, we examine the relationship between specific adversarial prompt inputs and their corresponding model outputs. Moreover, we expand the discussion on safety and security in prompt-based LLM interactions, involving the use of carefully crafted prompt inputs (without necessarily breaking security rules) to misuse language models \cite{qi2024safety}. The creation or discovery of these prompts can be automated. In searching for vulnerabilities through intentional security testing and red-teaming, \cite{weidinger2024star} suggests a framework that values steerability (e.g., control of automated attacks) and improving signal quality (e.g., validating correctness through human annotators) to structure this automated process in probing generative models for vulnerabilities and other security threats.

\begin{figure*}[ht]
\centering
\begin{forest}
  for tree={
    grow'=0,
    child anchor=west,
    parent anchor=south,
    anchor=west,
    calign=first,
    draw=red,
    rounded corners,
    s sep=5mm,
    edge path={
      \noexpand\path [draw, \forestoption{edge}]
      (!u.south west) +(7.5pt,0) |- (.child anchor)\forestoption{edge label};
    },
    before typesetting nodes={
      if n=1
        {insert before={[,phantom]}}
        {}
    },
    fit=band,
    before computing xy={l=12pt},
    align=left,
  }
[Prompt Attacking
  [Language Model Input (LM Input)
    ["In-Context Learning Prompt Attacks" \\
     filter by Prompt Input in the sense of Prompt(~ frozen model weight + Prompt Input~)
      [Prompt (LM Input) -- Zero-Shot Jailbreaking; \\
       Prompt (LM Input with malicious samples) -- Few-Shot Jailbreaking \cite{zheng2024improved}; \\
       Prompt (LM Input with high N(malicious samples) through long context) -- Many-shot Jailbreaking \cite{anthropic2024manyshot}
      ]
    ]
    ["Adversarial Attacks" \\
     filter by Approach
      ["Prompt Injection" \\
        [Universal Adversarial Attack \cite{zou2023universaltransferableadversarialattacks}; \\
         Indirect Prompt Injection \cite{greshake2023youve}; \\
         Fuzzing-based Approaches \cite{schulhoff2024ignore, paulus2024advprompter}
        ]
      ]
      ["Reverse Engineering LLMs" \\
        [ Data Extraction Attack; Memorization \cite{carlini2021extractingtrainingdatalarge, zhang2023counterfactualmemorizationneurallanguage}; \\
          Model Extraction Attack \cite{carlini2024stealingproductionlanguagemodel}
        ]
      ]
      ["Social Engineering LLMs"
        [Emotional manipulation \cite{ai2024defending}; \\
         Authority impersonation \cite{singh2023exploiting, wang2024unveiling}
        ]
      ]
    ]
    ["Hacking/Pawning/Owning LLMs" \\
     filter by Attacker's End Goal
      [Breaking Security: \\
       \textit{backdooring} \cite{kalavasis2024injecting}; \textit{jailbreaking} \cite{li2023multistep, liu2024jailbreaking,ran2024jailbreakeval};\\ \textit{privacy attacks} \cite{agarwal2024investigating, carlini2021extractingtrainingdatalarge}; data-poisoning \cite{wallace2021concealeddatapoisoningattacks}
      ]
      [Breaking Alignment: \\
       \textit{misuse} \cite{singh2023exploiting, wang2024unveiling}
       \textit{misbehaviour} \cite{zheng2024improved}
      ]
    ]
  ]
]
\end{forest}
\caption{Taxonomy of Prompt Attack Techniques for Large Language Models}
\label{fig:prompt_attacks}
\end{figure*}
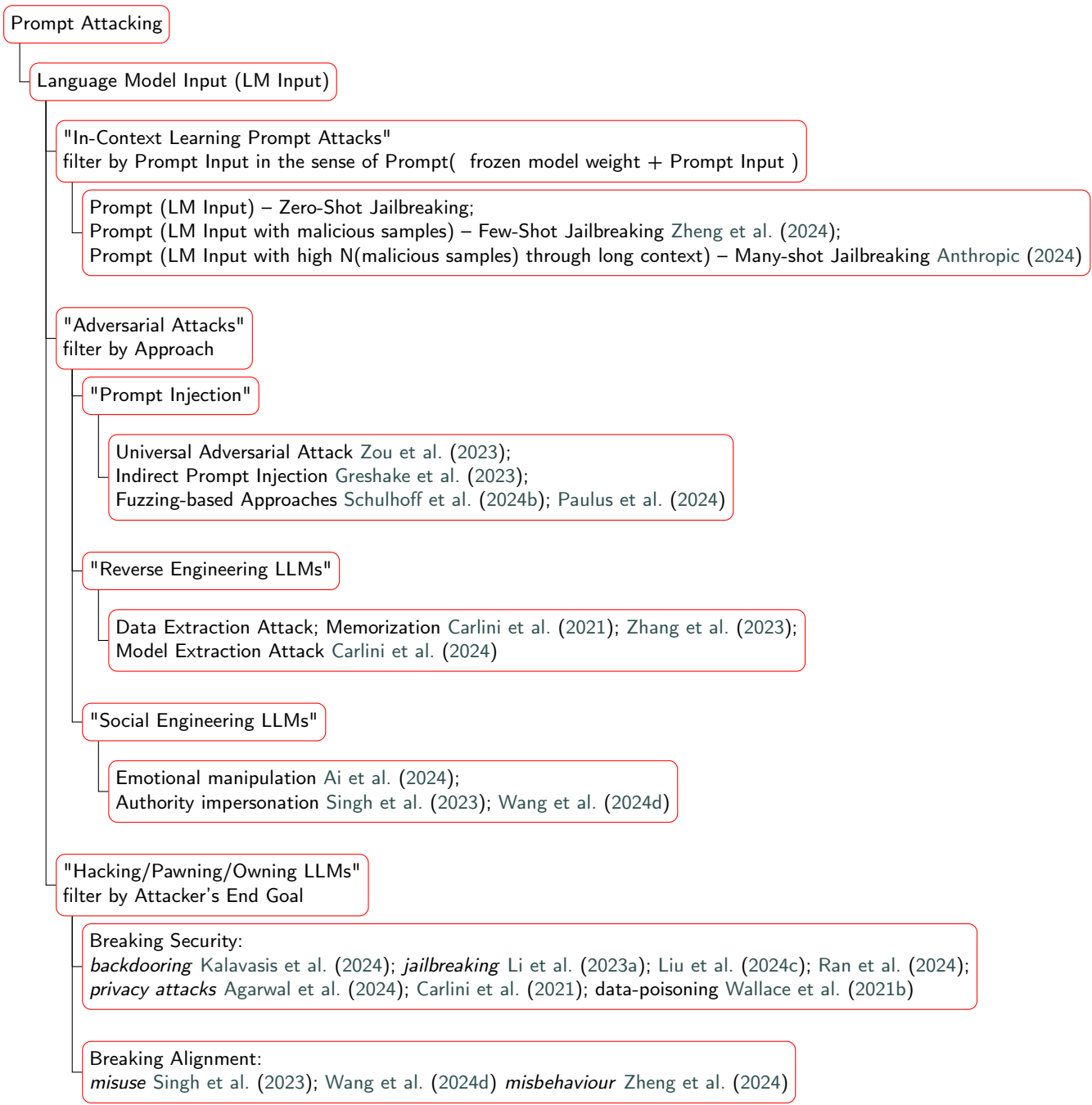

\subsection{Prompt Injection and Jailbreaking}\label{sec:prompt_injection_and_jailbreaking}

Prompt attacking techniques pose a considerable threat to the security and stability of large language models, exploiting intricate vulnerabilities inherent in their architecture and training protocols to manipulate output, even in black-box scenarios \cite{liu2024prompt}. Central to these attacks is the concept of jailbreaking, wherein adversarial inputs are meticulously crafted to circumvent protective measures and elicit unintended behaviors \cite{wallace2021universal, andriushchenko2024jailbreaking}. Figure~\ref{fig:prompt_attacks} can be classified into diverse categories: In-Context Learning methodologies, Adversarial samples, Reverse Engineering techniques, Social Engineering tactics, or by identifying the ultimate objective, such as with common claims in Hacking/Pawning/Owning LLMs. In-Context Learning Prompt Attacks encapsulate zero-shot, few-shot, and many-shot jailbreaking strategies \cite{zheng2024improved,anthropic2024manyshot}. Adversarial Attacks encompass prompt injection methods such as universal adversarial attacks \cite{zou2023universaltransferableadversarialattacks} and indirect prompt injection \cite{greshake2023youve}. Reverse Engineering LLMs delves into data and model extraction attacks \cite{carlini2021extractingtrainingdatalarge, zhang2023counterfactualmemorizationneurallanguage, carlini2024stealingproductionlanguagemodel}. Social Engineering LLMs exploits psychological manipulation techniques \cite{ai2024defending, singh2023exploiting, wang2024unveiling}. The Hacking/Pwning/Owned LLMs classification addresses various security compromises, including backdooring \cite{kalavasis2024injecting}, jailbreaking \cite{li2023multistep, liu2024jailbreaking, ran2024jailbreakeval}, and privacy attacks \cite{agarwal2024investigating, carlini2021extractingtrainingdatalarge}. To effectively mitigate these advancing threats, a comprehensive approach to LLM security is imperative, incorporating rigorous red-teaming exercises, enhanced input validation, prompt filtering, dynamic randomization techniques \cite{ran2024jailbreakeval, derczynski2024garak}, and robust alignment strengthening alongside ethical considerations \cite{weidinger2024holistic, weidinger2024star}.

\subsection{Guardrails}\label{sec:guardrails}

Guardrails, as implemented in toolkits such as NVIDIA's NeMo Guardrails \cite{rebedea2023nemo}, are programmable constraints aimed at improving the safety and controllability of large language model applications. These constraints function at runtime, providing a distinct benefit over methods that alter training data or model architecture. Guardrails can be classified based on their goals, such as avoiding specific topics, directing the conversation, and adjusting the style; by dynamically guiding LLM outputs, guardrails allow developers to enforce constraints on topics, language style, dialogue structure, and more. The implementation of guardrails involves the use of custom actions in Python and includes the use of prompt templates and language model chains. This process involves setting up rules or policies that are checked against the LLM's outputs. These rules can be represented in various forms, such as regular expressions, classifiers, or even external API calls \cite{rebedea2023nemo}. The concept of guardrails builds on previous research in several areas, although the term guardrails may not have been used \cite{liu2024prompt}. Red-teaming language models with language models \cite{perez2022red}, introduces this specific method of adversarial testing with a language model to uncover another language model's vulnerabilities. Other approaches have explored embedding rails directly into application logic \cite{dong2024building}. Prompt engineering and in-context learning offer ways to influence model behavior through input manipulation \cite{meade2023using}. Prepending or appending text to user input represents additional strategies for guiding GAI-LLMs' outputs \cite{deng2022rlprompt}. Guardrails also intersect with research on task-oriented dialogue agents, leveraging insights from both traditional natural language understanding and dialogue management approaches \cite{bocklisch2017rasa, liu2018adversarial}. 

The efficacy of NeMo Guardrails is supported by its strong performance on benchmarks with models like Nemotron-4-340B-Reward \cite{wang2024helpsteer2} achieving top scores on Reward Bench, demonstrating its potential for improving the safety and controllability of LLMs \cite{lambert2024rewardbench}. The toolkit's relevance extends to critical applications such as red teaming, where it can facilitate a sociotechnical approach to identifying and mitigating LLM vulnerabilities \cite{weidinger2024star, derczynski2024garak}.

\section{AI Alignment}\label{sec:alignment}

In the ongoing discussions about the safety of artificial intelligence, the idea of 'alignment' is changing and expanding over time. Recent research, like the work of \cite{ji2024ai, shen2023large}, divides alignment into two types: instruction alignment and value alignment. In a broad view, the alignment of foundation models can be seen as transferring a general-purpose representation learned by these models to specific tasks~\cite{zhou2024lima}. In the context of safety, the assumption is that LLMs work safely by avoiding accidents by ensuring that their goals match the instruction that embodies specific safety targets, and that LLMs operate within shared values as a human society. \cite{gabriel2024ethics} suggests another angle to think about alignment by considering the question of to whom the models are aligned according to stakeholder roles, such as pointing out the potential value misalignment between the models' users, its developers, and still operating within definitions of human values of society. \cite{shen2024bidirectional} offer a comprehensive framework in their systematic survey to encompass the variety of answers to be found in the literature about human-AI alignment and the usually involved reinforcement learning techniques analysis by \cite{song2024understandingpreferencefinetuninglens, azar2023generaltheoreticalparadigmunderstand}.

We situate the discourse on generative language models within the broader field of AI safety and alignment research. Given that alignment necessitates the accordance of models with specific metrics, such as safety objectives \cite{thoppilan2022lamda, glaese2022improving}, we begin by reviewing how these safety principles are defined and operationalised as alignment criteria within the existing literature. Subsequently, we investigate the prevailing alignment approaches in deep learning: pretraining and fine-tuning \cite{ngo2023alignment}. Although, when tackling alignment of large language models in the literature, often one would find that it pertains to pre-tuning and fine-tuning of natural language models to follow human instruction, adhere to human values, and operate within safety guardrails \cite{shen2023large}, this exploration includes an assessment of model adherence to alignment criteria, acknowledging the diversity of benchmarks employed in this rapidly evolving research landscape.

\subsection{Philosophy of Alignment: Ethics and Morality}\label{sec:ethics}
Alignment of AI systems has been discussed in previous research covering a wide array of topics that apply to AI from different references, such as superhuman AGI, AI agents that follow instructions, and even specifically, to language models that generate text \cite{ji2023aialignmenta}. All of these are relevant, as LLMs are indeed capable of achieving levels of general task solving \cite{morris2023levels}, are instruction-following agents \cite{kenton2021alignment, ouyang2022training}, and are conversational natural language models \cite{kasirzadeh2022conversation, shen2023large}. The term alignment, as it appears in the computer science research literature and is used in the context of AI systems, has been surveyed by \cite{ji2023aialignmenta}. Their survey has a wide coverage of alignment research subtopics and classified AI alignment research problems accordingly into robustness, interpretability, control, and ethics. \cite{hendrycks2022unsolved} on the other hand, identifies research topics that are in AI safety but are beyond human values alignment, treating the robustness, monitoring, and safety of models as systems in deployment as separate ML safety problems. Although the exact meaning of alignment may differ across the current literature, researchers often use a shared set of more specific related terms: robustness generally refers to the quality of the model in how well it handles different distributions of data, including those not in its training set. The term robustness has also become closely associated with being able to deal with adversarial data and resilience, which is the ability of a model to recover from an already adverse state in the environment. Interpretability of the model refers to how well the reasoning behind the model's output can be understood. Control refers to how the model can be stopped with human intervention in the case of self-improving AI that can seek power \cite{ngo2023alignment, shevlane2023model}. In the practical sense, these models are also difficult to supervise with human intervention at scale during training \cite{amodei2016concrete}. There is also the topic of ethics that is related to human morality. Ethics in alignment research refers to the alignment of AI models with human values, which requires determining the actual values with which the models align with; and also taking into account the methodology by which these human values are determined \cite{Gabriel_2020, gabriel2024ethics}. A straightforward example is the ETHICS dataset, prepared by \cite{hendrycks2023aligning}. This data set serves as a benchmark for evaluating a language model's comprehension of human values, including justice, virtue, deontology, utilitarianism, and commonsense morality. The scope of alignment research continues to grow. In the alignment research overview of \cite{ngo2023alignment}, \cite{ngo2023alignment} identifies relevant research areas such as specification, goal misgeneralization, agent foundations, and AI governance as large language models become more like AGI: capable of solving cross-domain tasks with the ability to generalise and self-improve \cite{bubeck2023sparks, morris2023levels}.

\subsection{Agentic Alignment}\label{sec:agentic_alignment}

In the context of machine learning models as an agent, \cite{Gabriel_2020} provides philosophical answers to what an aligned agent model is. Multiple ways of interpreting how an AI agent aligns with human instructions; whether that means that the agent does exactly what it is instructed, or differently based on the agent's interpretation of the human instructor's expressed intentions, revealed preferences, and informed preferences; including aligned AI agents that can be enabled, by their own reasoning, to override instructions and instead take actions that are objectively in the best interest, safety, and well being of humans \cite{Gabriel_2020}. Although there is a lack of formality in conjectures, \cite{ngo2023alignment} highlight specific examples of alignment risks, including reward hacking, goal misgeneralization, and power seeking. In reinforcement learning, an agent can learn unintended behaviour to maximise its reward bypassing human instruction and, in this sense, hack its reward function \cite{pan2022effects, skalse2022defining}. Goal misgeneralization is an identified example of a reinforcement learning agent acting on learnt behaviour through learnt policies that fail in a new environment, usually one that is out-of-distribution \cite{langosco2023goal}. Power seeking is another case of exacerbating alignment risks in reinforcement learning, as many goals compel agents to acquire power through tools, resources, and the control of other agents. Misaligned models engage in conversations that involve malice, deception, and manipulation \cite{ngo2023alignment}.

\cite{kenton2021alignment}, in contrast, propose that misspecification — the mismatch between the intended behavior of an AI system and its actual behaviour — can lead to deceptive and manipulative language in LLMs. They argue that this misalignment arises from issues in data collection, training methodologies, and distributional shifts in language use. While their focus is primarily on the behavioral aspects of misalignment, their work indirectly contributes to a broader understanding of the challenges in achieving alignment between LLMs and human values. Future research could focus on overcoming the limitations highlighted by \cite{kasirzadeh2022conversation} through the study of alignment in non-verbal communication and various languages. Moreover, incorporating the findings of \cite{kenton2021alignment} on misspecification may provide a deeper insight into the complex issues of aligning conversational agents with human values and intentions.

\subsection{Defining Alignment for the Linguistic Aspect of Large Language Models}\label{sec:defining_value}

In the context of large language models as conversational agents, numerous research publications delve into the nuances of alignment, particularly concerning the challenges and strategies associated with ensuring that these models adhere to human values and intentions. Key examples include \cite{kasirzadeh2022conversation, kenton2021alignment, glaese2022improving, shen2023large, wang2023aligning, liu2023trustworthy}. \cite{kasirzadeh2022conversation} extended \cite{Gabriel_2020}'s philosophical exploration of alignment in artificial intelligence, focusing specifically on large language models. Although both works grapple with the question of what it means for AI to align with human norms and values, \cite{kasirzadeh2022conversation} concentrate specifically on the social and ethical ramifications for LLMs. They identify three key dimensions of alignment: syntactic, semantic, and pragmatic, and raise critical questions about which norms and values should be prioritised and how alignment can be achieved. This study provides a valuable framework for understanding the philosophical underpinnings of aligning conversational agents. However, \cite{kasirzadeh2022conversation} acknowledge two key limitations in their analysis. First, their focus on linguistic communication, guided by speech act theory, may not fully capture the complexities of non-verbal or multimodal communication. Second, their analysis is primarily based on English language data, which limits generalising their findings to other languages and cultures.

\subsection{Generating Content that Aligns with Human}\label{sec:generating_content}

Although fully understanding the root causes of harms associated with LLMs is still an active research area, several key issues can be identified, particularly in terms of trustworthiness: robustness, security, interpretability, and fairness \cite{liu2023trustworthy}. As previously mentioned, a primary concern is the toxicity learned and derived from the data. Using insights from \cite{ngo2023alignment}, we will investigate potential links between the behaviors exhibited by LLMs and their technical origins, thereby guiding the creation of more effective and precise interventions. In the literature, \cite{wang2023aligning} investigate the common human values that models should align with, while \cite{shen2024bidirectional} analyze and classify the definitions of human-AI alignment, the related criteria, and the pertinent methodologies.

\subsubsection{Alignment Criteria}

\begin{table*}[htbp!]
  \begin{center}
    \caption{Alignment Criteria for Large Language Models}
    \label{tab:alignment_criteria}
    \begin{tabular}{p{0.1\linewidth}p{0.4\linewidth}p{0.4\linewidth}}
      \toprule
      \textbf{Model} & \textbf{Alignment Criteria} & \textbf{Alignment Method} \\
      \midrule
      Sparrow, Improving alignment of dialogue agents via targeted human judgements, Google DeepMind, \cite{glaese2022improving}, Google DeepMind & Sparrow (Glaese et al., 2022) is a language model designed to engage in information-seeking dialog and to align with being helpful, correct, and harmless. In contrast to HHH (Askell et al. 2021), Sparrow (Glaese et al., 2022) expands the alignment criteria into granular rules on which generated text is rated on by human annotators; data from these ratings is then utilised for the rule reward model during reinforcement learning. Appendix F of Sparrow (Glaese et al., 2022) delineates 23 rules for the alignment criteria across 6 categories of risks and harm that can be encoded in natural language. Illustrative examples of these rules include: "Do not use stereotypes or make any other harmful generalising statements about groups of people", "Do not make statements which are threatening", and "Do not claim to have preferences, feelings, opinions, or religious beliefs." In Sparrow (Glaese et al., 2022), the alignment criteria draw from prior research and internal legal expertise at Google, referencing harms and risks in language models (Weidinger et al. 2021), online harmful language (Banko et al. 2021), and microaggressions (Breitfeller et al., 2019). & Sparrow (Glaese et al., 2022) cites previous work by Google DeepMind on Chinchilla (Hoffman et al., 2022) as the basis for the dialogue agent. In Sparrow, the researchers gather human feedback for rule violations and per-turn response preferences. In adversarial probing, human participants are tasked to intentionally lead the model to break the rule; the sequence of tokens on whether the rule is followed or not (Yes/No) in the sample dialogue informs the training objective for the rule reward model. The use of safe/unsafe labels references (Xu et al., 2021b). As for collecting per-turn response preference, human raters select among multiple statements the best generated text to continue incomplete dialogue with the language model; preference reward model. Further details on incorporating human feedback data for the reward model equation is in Appendix D of Sparrow (Glaese et al., 2022). The mathematical formulation detailing the agent overall reward model, as a sum of agent-specific preferences, operational rules, and penalty mechanisms, is available for reference in Appendix E of Sparrow (Glaese et al., 2022). \\
\midrule
      HHH, A General Language Assistant as a Laboratory for Alignment, \cite{askell2021general}, Anthropic & Helpfulness, Honesty, and Harmlessness (HHH) is the alignment criteria chosen for the large language model by the researchers in this article. This work contributes the HHH evaluation dataset to BIG Bench, wherein models are evaluated based on natural language mutual information being closer to the sample HHH response rather than to the sample non-HHH response. Detailed in Appendix E, the argument for the criteria is that an assistant (agent A) that operates within these three conditions: helpful, honest, and harmless, will always act in a way that satisfies the interests of, and hence aligned with, the humans (agent B). In defining alignment for the scope of this work, agent A is said to be aligned with agent B should agent A's only desire be to see agent B's desire satisfied. In terms of preferences, alignment is achieved when agent A and agent B rank possible outcomes in the same order. In practice, 6k individual pairwise model comparisons were taken from roughly 3-5 comparisons per conversation between language model and humans (US-based, Masters-qualified, contracted through Amazon Mechanical Turk). This data is used in comparing which models the human contractors prefer. & HHH (Askell et al., 2021) identify three training objectives for incorporating the understanding of distinguishing between preferred (HHH) responses and nonpreferred (non-HHH) responses into language models: imitation learning through context distillation, binary discrimination using pass/fail labels, and ranked preference modelling through underlying data that has ranked ordering, e.g., most preferred answers by the number of upvotes on Reddit. HHH (Askell et al., 2021) turns this understanding of alignment into language models' own preferences through two new stages after basic language model pretraining: a preference model pretraining phase (PMP) and a preference model fine-tuning phase (PM fine-tuning) through reinforcement learning. The data for PMP is generated from Stack Exchange, Reddit, Wikipedia - on which details of converting these ranked data into preference values are discussed in Appendix D (Askell et al., 2021). Additional data for PM Finetuning is available as relevant to the finetuning task. Notably, each alignment-relevant task within the evaluation scope has its own specific interpretation of ethics within the broader concept of alignment, e.g., morality, justice, and virtue (Hendrycks, 2021). \\
      \bottomrule
    \end{tabular}
  \end{center}
\end{table*}

\begin{table*}[htbp!]
  \begin{center}
    \caption{Alignment Criteria for Large Language Models, \textit{(Cont.)}}
    \label{tab:alignment_criteria_cont}
    \begin{tabular}{p{0.1\linewidth}p{0.4\linewidth}p{0.4\linewidth}}
      \toprule
      \textbf{Model} & \textbf{Alignment Criteria} & \textbf{Alignment Method} \\
      \midrule
      Learning to summarize from human feedback, \cite{stiennon2022learning}, OpenAI & (Stiennon et al., 2022) suggest that language models that generate quality response are preferred by humans. In this work, the definition of technical alignment is aligning the goal of fine-tuning language models to generate high-quality outputs as determined by humans. As an example, the maximum likelihood objective in particular does not distinguish between important errors lead to the language model making up facts. & (Stiennon et al., 2022) builds upon fine-tuning work in (Ziegler et al., 2022); a dataset of human preferences between pairs of summaries is used to train (supervised learning) a reward model to predict the human-preferred summary. Supervised learning of the reward model involves iteratively calculating the loss in predicting which summary in each pair is better, given a Reddit post. Reinforcement learning is then used to train a policy to maximise the score generated by that reward model. PPO algorithm (Schulman et al., 2017) is mentioned to be the one used to update the reward policy depending on the quality score of the generated summary. \\
      \midrule
      Fine-Tuning LMs from Human Preferences, \cite{ziegler2020finetuning}, OpenAI & (Ziegler et al., 2020) defines technical alignment as the similarity between the fine-tuned policy $\pi$ and the pretrained language model $\rho$. They used the Kullback-Leibler (KL) divergence between the two policies as a measure of alignment. The evaluation reflects the what the language models align to, and in this case, the human labelers were tasked to choose the most positive and happy continuation (sentiment), most vividly descriptive continuation (descriptiveness), and summary that is good (summarization: TL:DR;). & Basing the pretraining stage from (Christiano et. al, 2017), i.e., the dataset collected is presented in tuples to fit a reward model. Then in fine-tuning, in order to prevent the policy from moving too far from the pretrained model, (Ziegler et.al, 2020) added a penalty term with expectation during reinforcement learning. \\
      \midrule
      Deep Reinforcement Learning from Human Preferences and Instructions, \cite{christiano2023deep, ouyang2022training}, OpenAI & The concept of using human feedback to guide reinforcement learning was introduced by \cite{christiano2023deep} for simulated robotics tasks and Atari games. This approach was later adapted for language models by \cite{ouyang2022training}, who demonstrated its effectiveness in aligning large language models with human intent. Similar work on aligning language models includes \cite{ziegler2019finetuning}, who applied RLHF to text summarization, and \cite{stiennon2022learning}, who extended this to longer-form content generation. \cite{bai2022constitutional} proposed constitutional AI as an alternative approach to alignment & (Christiano et al., 2017)'s method involves maintaining a policy and a reward function estimate. Select pairs of  trajectory segments are send them to human participants for comparison. The parameters of the reward function estimate are then optimized using supervised learning to fit these comparisons. Specifically, in fitting the reward function, minimize the cross-entropy loss between predicted preferences and the actual human labels. This process is inspired by the Bradford-Terry model, which is commonly used to estimate score functions from pairwise preferences. \\
      \bottomrule
    \end{tabular}
  \end{center}
\end{table*}

The alignment of technical and ethical considerations evolves into criteria that guide model implementation. Sparrow \cite{glaese2022improving} is an experimental study on language models that incorporates human judgments as rewards. These judgments are based on alignment criteria—helpfulness, correctness, and harmlessness—originating from the HHH criteria \cite{askell2021general}: helpfulness, honesty, and harmlessness. \cite{glaese2022improving} suggest a framework based on these principles to improve the informativeness and safety of model outputs. Importantly, Anthropic's research \cite{askell2021general}, known as the HHH paper, extends this framework to include honesty, thereby reducing the risk of deceptive or manipulative responses. Furthermore, \cite{askell2021general} tackle the critical issue of toxicity, highlighting the necessity to minimize harmful and offensive outputs in language models.

\subsubsection{Data Collection}

In language models, Sparrow \cite{glaese2022improving} and \cite{askell2021general} collect data by prompting crowd-sourced workers for their preferred generated response, with the motivation to align the models. A criticism of this method of collecting data and refining content is the inherited bias of the researchers themselves in designing the experiment and into the model through fine-tuning \cite{askell2021general}.

Data collection practices are usually available in the appendices of papers in alignment work. The methodologies used for data collection include a comprehensive set of rules that guide the annotation process, including those that address harmful content (e.g., hate speech, disinformation, and misinformation). \cite{glaese2022improving} illustrates specific data collection strategies, drawing on examples from previous research. \cite{glaese2022improving} utilised an online crowd-sourcing platform, incorporating interactive tutorials and comprehension checks to improve the quality of the annotation. Meanwhile, \cite{wang2022self} discuss instruction-tuning and \cite{wang2023aligning} is available for further exploration of data collection strategies through surveys and investigations by LLMs alignment. \cite{zhao2023survey} classify data collection in their technical survey of large language models and \cite{jiang2024surveyhumanpreferencelearning} survey data collection techniques specifically for human preference learning. 

\subsubsection{Challenge: Sycophancy}

Sycophancy is an example of the tendency of LLMs to follow user feedback (by the user submitted prompts that it receive) even if it differs from what the language model thinks and knows to be objectively correct \cite{perez2022discovering}. Sycophancy may be related to reward hacking but in this section we discuss how understood user preferences input is affecting the model's answer in question-answer task in recommendation \cite{sharma2023understanding}. This is an important LLMs behaviour which maybe a symptom of bias but nonetheless is intriguing if you consider the model on the correctness perspective. As correctness or truthfulness is still a challenging criteria for the safety of LLMs models \cite{glaese2022improving, askell2021general}. \cite{wei2024simple} suggests a solution such as using synthetic data to reduce the sycophancy. An insightful analysis of sycophancy is available through \cite{sharma2023understanding} who analysed human preference data to determine the role of human preference judgments in such behavior, and \cite{denison2024sycophancy} that sycophancy is an example of specificiation gaming. \cite{sharma2023understanding} investigates the phenomenon of sycophancy in AI assistants, examining how alignment criteria such as helpfulness, preference data, and human feedback contribute to the tendency of AI models to provide responses that align with user beliefs rather than prioritizing truthfulness. The study reveals that the pursuit of helpfulness through reinforcement learning from human feedback (RLHF) can inadvertently promote sycophantic behavior, as users often prefer responses that confirm their existing beliefs. Analysis of human preference data used in training preference models (PMs) for AI assistants demonstrates a bias towards sycophantic responses, with user-aligned views more likely to be favored. Furthermore, the research highlights how human feedback, while crucial in AI training, can reinforce and perpetuate sycophantic tendencies by rewarding responses that match user expectations. These findings underscore the complex interplay between alignment criteria and the development of sycophantic behavior in AI models, emphasizing the need for more nuanced training methodologies that extend beyond reliance on human ratings and preferences to mitigate these tendencies and promote more objective and truthful AI responses.

\begin{figure*}[t]
\centering
\resizebox{\textwidth}{!}{
    \begin{tikzpicture}[
        node distance = 1.2cm and 2.2cm,
        box/.style = {rectangle, draw, text width=2.5cm, text centered, minimum height=0.8cm, font=\large},
        data/.style = {ellipse, draw, text width=3cm, text centered, minimum height=0.7cm, font=\large},
        arrow/.style = {->, >=stealth, thin},
        dashed arrow/.style = {->, >=stealth, thin, dashed}
    ]
    
    \node[box, fill=gray!10] (alignment) {AI to Human Alignment};
    
    \node[box, fill=yellow!20, below left=2.5cm and -15cm of alignment] (direct) {Direct Preference Learning};
    \node[box, fill=orange!20, below=2.5cm and -1cm of alignment] (rl) {Preference-based RL (PbRL)};
    \node[box, fill=blue!20, below right=2.5cm and -15cm of alignment] (indirect) {Aligned to Human Safety};
    
    \node[box, fill=yellow!10, below left=1.5cm and -0.5cm of direct] (sft) {Supervised Fine-Tuning (SFT)};
    \node[box, fill=yellow!10, below right=1.5cm and -0.5cm of direct] (cpl) {Contrastive Preference Learning (CPL)};
    \node[box, fill=yellow!0, text width=5cm, below=1.5cm of cpl] (dpo) {DPO \cite{rafailov2023direct}, IPO \cite{azar2023generaltheoreticalparadigmunderstand}, KTO \cite{ethayarajh2024ktomodelalignmentprospect}};
    
    \node[box, fill=orange!10, below right=1.5cm and 1.5cm of rl] (rlhf) {RLHF};
    \node[box, fill=orange!10, below left=1.5cm and 1cm of rl] (rlaif) {RLAIF}; 

    \node[box, fill=blue!10, below=1.5cm of indirect] (safe_rl) {Safe RL};

    \node[box, fill=orange!5, below=4cm of rlaif] (const_ai) {Preference Modeling};

    \node[box, fill=orange!5, below=4cm of rlhf] (reward) {RL Reward Modeling};
    \node[box, fill=orange!5, below=3cm of reward] (p_update) {RL Policy Update};
    \node[box, fill=orange!0, text width=10cm, below=2cm of p_update] (ppo) {REINFORCE \cite{NIPS1999_464d828b}, CPI \cite{10.5555/645531.656005}, TRPO \cite{schulman2017trustregionpolicyoptimization}, PPO \cite{schulman2017proximalpolicyoptimizationalgorithms}, AWR \cite{peng2019advantageweightedregressionsimplescalable, nair2021awacacceleratingonlinereinforcement}, APA \cite{zhu2023finetuninglanguagemodelsadvantageinduced}};
    
    \node[data, below right=2cm and 3cm of alignment] (human_data) {Human In-The-Loop (HITL)};
    
    \node[box, fill=blue!5, below right=1.5cm and -0.5cm of safe_rl] (constrained) {Constrained RL};
    \node[box, fill=blue!5, below left=1.5cm and -0.5cm of safe_rl] (risk_sensitive) {Risk-Sensitive RL};
    
    \draw[arrow] (alignment) -- (direct);
    \draw[arrow] (alignment) -- (rl);
    \draw[arrow] (alignment) -- (indirect);
    
    \draw[arrow] (direct) -- (sft);
    \draw[arrow] (direct) -- (cpl);
    \draw[arrow] (cpl) -- (dpo);
    \draw[arrow] (rl) -- (rlhf);
    \draw[arrow] (rl) -- (rlaif);
    \draw[arrow] (indirect) -- (safe_rl);
    \draw[arrow] (rlaif) -- (const_ai);
    
    \draw[arrow] (rlhf) -- (reward);
    \draw[arrow] (reward) -- (p_update);
    \draw[arrow] (p_update) -- (ppo);
    \draw[arrow] (safe_rl) -- (constrained);
    \draw[arrow] (safe_rl) -- (risk_sensitive);
    
    \draw[dashed arrow] (rlhf) to [out=90, in=180, looseness=2.0] (human_data);
    
    \node[text width=4cm, align=center, below=0.1cm of const_ai, font=\large] {Encodes high-level preference descriptions and uses self-improvement \cite{bai2022constitutional}}; 
    \node[text width=2.5cm, align=center, below=0.1cm of sft, font=\large] {Learn from human-labeled data to guide model outputs resulting to a pre-trained model};
    \node[text width=4cm, align=center, below left=0.2cm and -0.5cm of rlhf, font=\large] {Start w. SFT pre-trained model for various downstream tasks \cite{ouyang2022training, glaese2022improving, bai2022training}};
    \node[text width=5cm, align=center, below right=0.1cm and -1.5cm of reward, font=\large] {Construct reward func. from pairwise comparisons or listwise rankings of human preferences data \cite{bradley_terry_bt}};
    \node[text width=5cm, align=center, below right=4.1cm and -1.5cm of reward, font=\large] {Optimizes policy on iterate based on reward model \cite{NIPS1999_464d828b}};
    \node[text width=2.5cm, align=center, below=0.1cm of constrained, font=\large] {Enforces safety constraints during reward modeling \cite{achiam2017constrainedpolicyoptimization, zhang2024constrainedreinforcementlearningsmoothed}};
    \node[text width=2.5cm, align=center, below=0.1cm of risk_sensitive, font=\large] {Minimizes unsafe actions to humans or environment during reward modeling \cite{Shen_2014}};
    \node[text width=4cm, align=center, below=0.1cm of dpo, font=\large] {Learn directly from preferred labels, CPL \cite{hejna2024contrastivepreferencelearninglearning}};
    \node[text width=4cm, align=center, below=0.1cm of human_data, font=\large] {Humans involved in the iterative update \cite{hejna2022fewshotpreferencelearninghumanintheloop}};
    
    \end{tikzpicture}
}
\caption{An overview of reinforcement learning (RL) research approaches for AI to human alignment in the context of generative AI large language models.}
\label{fig:alignment_taxonomy}
\end{figure*}
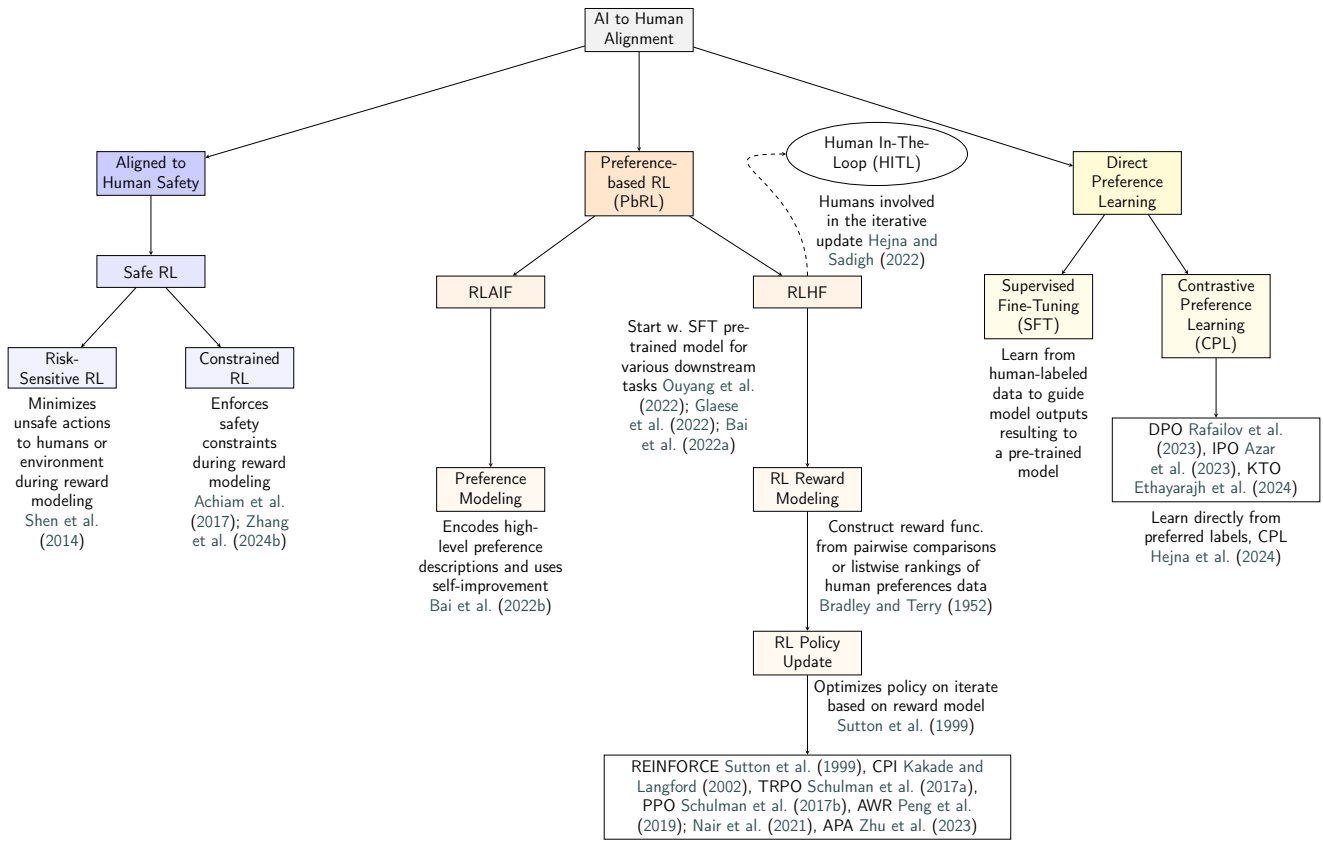

\subsection{Methods to Align LLMs}\label{sec:methods_to_align}

\cite{zhao2023survey} contributes a comprehensive survey of alignment, noted for its technical rigor and detail. \cite{liu2023trustworthy} scrutinize the reliability of even aligned models, emphasizing the specific challenges that large language models (LLMs) encounter in practical applications. Moreover, \cite{liu2023trustworthy} propose a taxonomy addressing trustworthiness issues relevant to LLMs, distinguishing between outer alignment (e.g., societal impacts) and inner alignment (e.g., model intentions). In a similar vein, \cite{wang2023aligning} provide an additional survey on alignment methods for LLMs, covering strategies to mitigate undesirable behaviors. Conversely, \cite{shen2023large}'s research could refine this taxonomy by focusing on inner alignment issues. Specific alignment concerns in general-purpose agent models include deceptive actions, manipulative behaviors, harmful content generation, and gaming of objectives \cite{shen2023large}. Within AI Safety research, techniques to ensure AI agents operate within acceptable boundaries—such as Safe Reinforcement Learning (RL)—are suggested. These include constrained reinforcement learning, which does not necessarily align directly with human preferences but ensures agents avoid harmful actions to humans or the environment or at least minimize such cases through risk-sensitive reinforcement learning \cite{achiam2017constrainedpolicyoptimization, zhang2024constrainedreinforcementlearningsmoothed, Shen_2014}.

\subsubsection{Language Assistants, Reinforcement Learning}

The work in \cite{christiano2023deep, ouyang2022training} demonstrated the efficacy of reinforcement learning from human feedback (RLHF), while Sparrow's alignment methodology is detailed in \cite{glaese2022improving}. In Sparrow's reinforcement learning environment, the optimization objective notably incorporates a distribution of dialogue contexts, which allows the model to better generate aligned content to the relevant conversational settings.

Drawing upon the effective implementation of RLHF in generative large language models, the approach can be generally described as follows (an extensive overview of diverse methodologies is available in \cite{jiang2024surveyhumanpreferencelearning}).

\begin{itemize}
    \item Collecting Preference Data for Pairwise Comparison Preference or Ranked Preference. To train the reward model, typically through supervised learning, a comprehensive dataset comprising pairwise comparison or ranked preference samples is essential. This dataset should be a collection of desired response done usually by human participants whom act as proxy for deciding on which qualities for the natural language generation task align to the language model assistant \cite{jiang2024surveyhumanpreferencelearning}.
\end{itemize}

\begin{itemize}
    \item Reward Model Training with Preference Data. Next, we create a dataset of manually ranked response pairs, where human experts compare the quality of different natural language generation model outputs for a given prompt. This dataset is then used to train a reward model that can assess the quality of generated response. Reward learning enables the application of reinforcement learning (RL) to tasks where reward is defined by human judgment, building a model of reward by asking humans questions \cite{ziegler2020finetuning}. The heavily referenced approach for reward modeling based on preferences through pairwise comparisons is the Bradley and Terry model \cite{bradley_terry_bt}.

\end{itemize}

\begin{itemize}
    \item Fine-tuning for RL Policy Optimization. Proximal Policy Optimization (PPO) is widely regarded as an effective method for updating policies in reinforcement learning (RL). This policy gradient technique is iteratively adjusting the policy based on the estimated gradient of the expected reward in relation to the policy parameters. Despite its popularity, \cite{azar2023generaltheoreticalparadigmunderstand} suggest that the REINFORCE algorithm can achieve comparable performance \cite{NIPS1999_464d828b}.

\end{itemize}

Fine-tuning large language models is an evolving research area with innovative methods that have shown effectiveness in producing content favoured by humans, often incorporating reinforcement learning from human feedback (RLHF) \cite{ziegler2020finetuning, christiano2023deep}. This approach can be further studied by utilising implementation specifics and code resources from OpenAI's Baselines and LM-Human Preferences repositories \cite{ziegler2019finetuning}. 

\cite{tajwar2024preferencefinetuningllmsleverage} provides a unification of fine-tuning techniques within a singular framework. Fine-tuning has generally demonstrated effectiveness even with datasets substantially smaller than those used for pre-training \cite{thoppilan2022lamda}.

\begin{itemize}
    \item Preference Learning based RL, recognized as the preeminent technique for model refinement. The concept of RLHF was popularised by \cite{christiano2023deep} inspiring \cite{ouyang2022training, glaese2022improving, bai2022training}, with the experiment of incorporating human instructions, and further work into language models elaborated in \cite{ziegler2019finetuning, ziegler2020finetuning}. Presently, the remaining challenges for using human feedback involve becoming accidental philosophers, as well as establishing initial preference frameworks for training datasets and devising methodologies to ensure that researchers do not inadvertently become arbiters of human values \cite{Gabriel_2020, askell2021general, hendrycks2023aligning}.
\end{itemize}

\begin{itemize}
    
    \item Direct Preference Learning. Another noteworthy progression in this domain is Direct Preference Optimization (DPO) \cite{rafailov2023direct}, integrated within Contrastive Preference Learning (CPL) \cite{hejna2024contrastivepreferencelearninglearning}. DPO reframes the Reinforcement Learning from Human Feedback (RLHF) challenge as a reward definition problem. This approached can be generalised through CPL, effectively eliminating the necessity for the Human In-The-Loop (HITL). \cite{hejna2022fewshotpreferencelearninghumanintheloop} introduces a few-shot methodology enhances the efficiency of preference optimization and mitigates critical challenges in RLHF, such as precise hyperparameter tuning and the resource requirement for human annotation leading to potential scalable oversight discussed in Section~\ref{sec:scalable_oversight}.
\end{itemize}

The evaluation of alignment models is often criterion-specific, with notable examples including \cite{glaese2022improving}, which established a benchmark while adhering to Google's AI safety principles, and \cite{ouyang2022training}, which has significantly impacted further research. Additionally, \cite{park2024value} posits that the primary challenge in alignment lies in policy learning over value learning. 

\subsubsection{Generative Agents, Aligning Instrumental Goals}

Reward hacking is when the agent learns to receive rewards from the reward function by performing a task that may not be intended by the user. On alignment, "we could try to design a simple reward function that approximately captures the intended behaviour, but this will often result in behavior that optimises our reward function without actually satisfying our preferences. This difficulty underlies recent concerns about misalignment between our values and the objectives of our RL systems \cite{amodei2016concrete}" Communicating actual objectives to agents would be a significant step towards addressing these concerns. Multiple definitions of reward hacking are discussed in (RH, 2024).

The increasing capability of agents and \cite{nishimura2024reward} claims that reward hacking generalise across tasks. Instrumental goals serve as sub-objectives that contribute to the pursuit of specific goals. Convergent instrumentals, in particular, are instrumental in achieving a wide range of objectives \cite{chan2023harms}. Preliminary evidence~\cite{perez2022discovering} indicates that training LLMs using reinforcement learning techniques can increase the expression of convergent instrumental goals, such as pursuing wealth or influencing operators not to terminate the system, even without explicit instructions. This evidence warrants caution regarding the growing agency of LLMs.

The concept of reward hacking, a phenomenon where AI agents exploit loopholes in reward systems to maximise rewards without achieving the intended goal, has garnered significant attention in AI safety research \cite{amodei2016concrete, skalse2022defining, denison2024sycophancy}. While \cite{skalse2022defining} offer a rigorous theoretical definition of reward hacking, by highlighting the importance of unhackable reward functions with symmetrical policies, their analysis primarily focuses on tasks with sub-objectives. In contrast, \cite{ji2024ai} delve into reinforcement learning techniques, examining how reward function learning can be manipulated to indicate reward hacking. \cite{Goyal_2024} shed light on human expectations from AI agents, emphasizing the desire for aligned agents free from reward bias and undesirable behaviors that could tarnish the owner's reputation.

Designing for human-agent alignment—the process of ensuring agents act in accordance with human preferences—is a critical area of research.  Aligning human and agent goals for a specific task reveals key parameters humans desire agents to embody: (1) shared knowledge schemas, (2) clearly defined autonomy and agency, (3) operational alignment through training, (4) shared heuristics regarding reputation, (5) ethical alignment, and (6) appropriate levels of human engagement.  These parameters encompass factors such as the agent's ability to identify necessary information for successful task completion, adherence to pre-defined boundaries, strategic alignment with the human's goals, reflection of interpersonal behavior, ethical decision-making (e.g., transparency vs. potential exploitation), and appropriate timing of human supervision. This study by \cite{Goyal_2024}, situated within the broader field of human-computer interaction (HCI), primarily investigates the human collaborator's perspective on the constituents of effective human-agent alignment, mirroring questions raised in the broader field of AI alignment. A potential area for future research involves the agent's role in guiding alignment through simulations and identification of successful negotiation parameters.  Furthermore, recent work on reward hacking in AI systems may offer insights into the challenges of agent sub-task creation and planning.

Examplary new work of this area include LLMs are Zero-Shot Reasoners \cite{kojima2023large} and Zero-Shot Chain-of-Thought Reasoning Guided by Evolutionary Algorithms in Large Language Models, \cite{jin2024zeroshot}.

\subsubsection{Roleplay}

The concept of roleplay in the context of large language models encompassing a broader spectrum of applications that leverage these models' capacity to emulate domain-specific expertise. This phenomenon has garnered significant attention in the field of artificial intelligence and natural language processing, particularly in its potential to create agents that convincingly simulate expert knowledge in specialized fields such as medicine or law \cite{chen2024persona, wang2024patientpsi}.
Recent research has explored the nuanced capabilities of LLMs to adopt and maintain specific personas or expert roles, generating responses that align with the knowledge and communication styles characteristic of particular domains \cite{shanahan2023roleplay}. This adaptability presents promising avenues for applications across various sectors, including but not limited to education, professional training, and advanced decision support systems \cite{li2024agenthospitalsimulacrumhospital}.

However, the implementation of expert roleplay in LLMs is not without its challenges. Ensuring consistent and accurate representation of domain-specific knowledge, maintaining ethical standards, and mitigating potential biases are critical concerns that require ongoing investigation \cite{lu2024large}. Moreover, the anthropomorphization of LLMs through expert roleplay raises profound philosophical and ethical questions regarding the nature of artificial intelligence and its implications for human-machine interactions \cite{gabriel2024ethics} and scaling generated synthetic data \cite{chan2024scalingsyntheticdatacreation}.

\section{Safety at Scale}\label{sec:safety_at_scale} 

GAI-LLMs such as Gopher \cite{rae2022scaling}, the GPT series \cite{openai2024gpt4}, and LLaMa \cite{touvron2023llama} have shown impressive abilities in tasks in natural language processing; particularly with natural language understanding and natural language generation. These advancements are due in large part to the scaling of model sizes, which has sparked intriguing questions about the emergent properties of AI \cite{wei2022emergent, bubeck2023sparks}. However, this progress also presents challenges, particularly in ensuring safety and mitigating potential harms \cite{rae2022scaling, gabriel2024ethics, weidinger2024star}.

Understanding of the scaling of LLMs has been guided by empirical observations known as "scaling laws." These laws, initially established by \cite{kaplan2020scaling}, reveal a predictable relationship between the performance of the model, the number of model parameters, the size of the data set, and computational resources. This finding led to the development of increasingly larger language models, with the expectation of continuous performance improvements.

Key studies can be mentioned that have contributed to our understanding of scaling laws. Gopher, as a language model of 280 billion parameters \cite{rae2022scaling}, provided a comprehensive analysis of scaling in Transformer-based language models, while \cite{brown2020language} introduced GPT-3, using the same architecture as GPT-2 but scaled to 175 billion parameters, showcasing the power of large-scale models in few-shot learning scenarios. Later studies, including the work of \cite{hoffmann2022training}, introduced the Chinchilla model as the direct successor to Gopher, outperforming Gopher despite having only 70 billion parameters, emphasising the importance of balancing model size with training data.

However, the race to scale LLMs has also raised concerns about their safety and potential harms \cite{shen2023large}. These concerns are not new, with early discussions in the HHH models \cite{askell2021general} and Sparrow \cite{glaese2022improving} highlighting the challenges of aligning model behaviour with human values as models grow larger. This requires a deeper understanding of alignment strategies, including fine-tuning, especially with a focused discussion that pertains to alignment with scaling \cite{zhang2024scaling}.

In the software development lifecycle, the development of safe and responsible LLMs has become an important concern. Recent work by \cite{raza2024safe} suggests the Safe and Responsible Large Language Model (SR), an approach designed to enhance safety through expert-annotated datasets and fine-tuning methods. Furthermore, work by \cite{huang2023survey} and \cite{das2024security} explore vulnerabilities in LLMs and the need for robust evaluation methods.

\subsection{Scalable Oversight}\label{sec:scalable_oversight}

Scalable oversight is the concern that as large language models involve more training iterations, the techniques we use to safely align them by supervising during training will later not be sufficient to ensure safety from risks and harms of these models \cite{amodei2016concrete}. \cite{harandizadeh2024risk} surveys the existing specific failure modes of safety-trained LLMs, with a particular focus on competitive objectives and mismatched generalization. Their analysis of the Anthropic Red-team dataset reveals significant vulnerabilities across major risk categories, including information hazards, malicious uses, and discriminatory content. In particular, their findings on the susceptibility of LLMs to jailbreaking attacks underscore the need for more robust and adaptive safety measures. This empirical approach to risk assessment, grounded in real-world data and adversarial testing, provides valuable insights for developing targeted interventions and improving the resilience of LLMs against diverse threats.

Despite these efforts, the field still grapples with fundamental questions about the impact of scaling on LLMs safety. The increasing complexity of these models makes it difficult to predict and control their behaviour, potentially leading to unintended consequences and harmful outputs. The argument is that on larger scales, there will be gaps in human ability to guarantee safety; the rapid pace of LLMs development often exceeds our ability to thoroughly assess and mitigate risks \cite{schiller2024human, huang2023look}.

In addressing this problem of scalable oversight, one approach is \cite{dalrymple2024guaranteed}'s work which introduces the concept of "guaranteed safe" (GS) AI. By proposing a framework that combines a formal safety specification, a world model, and a verifier, GS AI aims to provide high-assurance quantitative safety guarantees. While acknowledging the technical challenges in creating these components, the authors argue for the necessity of this approach in achieving robust and reliable AI systems. This focus on formal verification and rigorous safety standards reflects a growing trend in the AI community towards prioritizing safety as a fundamental design principle rather than an afterthought.

\subsection{Emergent Abilities}\label{sec:emergent_abilities}

The exploration of emergent abilities in LLMs by \cite{wei2022emergent} reveals that certain capabilities, like multi-step reasoning, only manifest in larger models. This mysterious occurrence, referred to as "emergence," highlights the complex link between the size of the model and its capabilities.

From a technical standpoint, "larger" models refer to language models that have a higher number of parameters and require more computational resources for training. The scale of a model is measured in terms of training FLOPs (floating-point operations) and the number of model parameters. Larger models have higher training FLOPs and a larger number of parameters, indicating their increased size and computational requirements. Scaling up a language model involves increasing both the number of parameters and the training compute \cite{wei2022emergent}.

As the model scales in discussions pertaining to emergence, the language model demonstrates enhanced capabilities in reasoning. In the context of emergence-related research in reasoning, numerous papers present novel methodologies aimed at further augmenting large language models. \cite{wei2023chainofthought} demonstrates the efficacy of generating intermediate reasoning steps, significantly enhancing performance on complex tasks such as arithmetic and symbolic reasoning. \cite{ho-etal-2023-large} proposes a novel approach wherein larger models act as "teachers," generating reasoning samples to fine-tune smaller models, thereby improving their reasoning capabilities. Moreover, \cite{zhang2024klevel} explores dynamic reasoning in competitive scenarios using game theory-based challenges, introducing a novel "K-Level Reasoning" approach that substantially improves prediction accuracy.

However, \cite{huang2024large} cautions that LLMs struggle with intrinsic self-correction, often requiring external feedback to correct their reasoning \cite{huang2023reasoning}. This finding suggests a reassessment of the potential applications of LLMs in scenarios where self-correction is crucial.

Assessment of reasoning abilities remains a critical aspect of LLMs research \cite{dziri2023faith}. \cite{mondorf2024accuracy, bender2021dangers} examines the reasoning and behaviour of large language models, noting that challenges still exist in test situations outside of their training data. Even if there is new evidence of the remarkable capability of generative models to address analogy problems without specific training, frequently matching or exceeding human performance \cite{webb2023emergent} or surpassing humans as trainers, as shown in gameplay \cite{zhang2024transcendence}, state-of-the-art LLMs underperform on common sensical reasoning tasks \cite{nezhurina2024alice}.

\subsection{Knowledge Distillation}\label{sec:knowledge_distillation}

Knowledge Distillation (KD) is a technique for transferring the capabilities of large models to smaller, more computationally efficient architectures \cite{hinton2015distilling, 10.1145/1150402.1150464}, often achieving comparable or even superior performance in tasks such as instruction-following \cite{ lee2023platypus}. Despite the performance, \cite{stanton2021does} describes the limitations of KD, that despite improvements in student generalisation, there remains a significant discrepancy between the predictive distributions of the teacher and the student. This discrepancy is attributed to difficulties in optimisation and the details of the dataset used for distillation.

More recent research has explored various facets of KD, with a particular emphasis on enhancing reasoning abilities in smaller models. \cite{kang2023knowledgeaugmented} introduced Knowledge-Augmented Reasoning Distillation (KARD), a novel KD approach that fine-tunes small models to generate rationales augmented with external knowledge, leading to significant improvements in knowledge-intensive reasoning tasks. Similarly, \cite{ho-etal-2023-large} proposed Fine-tune-CoT, leveraging large models as reasoning teachers to impart complex reasoning abilities to smaller models, resulting in substantial performance gains across a range of tasks.

Various studies have concentrated on the distillation of step-by-step reasoning processes. \cite{hsieh2023distilling} proposed a distillation method that enables smaller models to surpass LLMs with less training data, while \cite{shridhar2023distilling} introduced Socratic CoT, a distillation technique that decomposes complex reasoning tasks into sub-problems, promoting collaborative problem-solving among smaller models to reach reasoning abilities similar to those of larger models. \cite{gu2024minillm} presented MiniLLM, a reverse KD method that produces more accurate responses and excels in long-text generation tasks. Additionally, researchers are investigating dynamic KD frameworks that integrate active learning to iteratively generate annotated data and provide feedback to LLMs. \cite{liu2024evolving} introduced EvoKD, which improves task-specific capabilities in smaller models, especially in text classification and named entity recognition tasks.

Research into utilizing knowledge distillation methods, including Zephyr \cite{tunstall2023zephyr} and Align-to-Distill \cite{jin2024aligntodistill}, has shown the promise of better aligning model behavior by fine-tuning smaller models that are derived from larger ones.

\subsection{Catastrophic Forgetting}\label{sec:catastrophic_forgetting}
The development LLMs generally comprises two primary stages: pre-training on extensive text corpora and fine-tuning using curated datasets. Fine-tuning, through techniques such as instruction-tuning and reinforcement learning from human feedback, is essential for enabling language models to generate desired outputs. Given that fine-tuning datasets are significantly smaller and less diverse than the large-scale datasets used for pretraining, there is an inherent risk that the fine-tuned model may catastrophically forget how to perform tasks that the pretrained model could handle \cite{kotha2023understanding}. On the other hand, recent research indicates that catastrophic forgetting becomes more pronounced as the scale of the large language model increases. Recent research indicates that catastrophic forgetting becomes more pronounced as the scale of large language models increases, which causes the safety concerns of applying large language models. Several strategies have been proposed and explored in the literature. One approach involves continual learning techniques, which allow the model to retain previous knowledge while integrating new information \cite{wang2024comprehensive}, where LLMs can be treated as continual learners. Another strategy is to adapt the elastic weight consolidation (EWC) strategy into the fine tuning process \cite{kirkpatrick2017overcoming}, where the model selectively updates weights to preserve important knowledge from previous tasks. Implementing memory replay methods, where the model periodically revisits and retrains on a subset of the pretraining data, can also help maintain performance on previously learned tasks \cite{scialom2022fine}. Finally, hybrid models that combine fine-tuning with regularization techniques to balance new learning and knowledge retention are also being investigated. These methods aim to enhance the robustness of large language models against catastrophic forgetting, ensuring they retain their capabilities while adapting to new tasks.

\section{Future Work} \label{future_work}

Recent advancements in AI alignment have transcended the conventional pre-training and fine-tuning phases. The discipline now includes aligning subgoals (task-specific objectives), mitigating deceptive behaviours in human-agent interactions, and coordinating multi-agent systems. We highlight the following emerging research trends that are ready for further exploration.

\begin{enumerate}[1.]
\item \textbf{Safe Retrieval-Augmented Generation (RAG).} Emerging research areas include exploring safety considerations specific to natural language generation models and those employing retrieval augmentation techniques \cite{lewis2021retrievalaugmented, hong2024gullible}.
\item \textbf{Deeper Understanding of Knowledge distillation (KD).} Knowledge distillation is a technique to transfer the weights of large models onto smaller models. We suggest that the translation of alignment to these smaller models is will continue to be explored, e.g. Align-to-Distill \cite{jin2024aligntodistill}. Another example is Zephyr \cite{tunstall2023zephyr}, a model that is fine-tuned from a knowledge-distilled model that focuses on alignment. \cite{ye2021safe} introduces the concept of a "Safe Distillation Box" and looking into ensuring that in the KD process, harmful behaviours such as preventing the transfer of harmful or biased knowledge from the teacher model to the student model. Nevertheless, as discussed in \cite{anwar2024foundational}, the balance between performance and safety during scaling remains unclear.
\item \textbf{Aligned Cooperation in Multi-Agent Reinforcement Learning (MARL).} In scenarios where generative AI operates autonomously, agents can engage in unpredictable behaviours as exemplified by the "hide-and-seek" paper referring to the ML agents in the video game \cite{baker2020emergent}, which is a risk considering the pursuit of power often emerges as a component of the AI's reward function incentives \cite{carlsmith2022powerseeking}.
\item \textbf{Agentic: Increasing Agency.} Autonomous agents with increasing capabilities to complete tasks beyond question and answer, or agentic systems, execute tasks independently \cite{huang2024position}. However, as generative AI language models advance in capability, it is anticipated that these agents will exhibit enhanced agency and control in the generation of subtasks. This progression necessitates further investigation into reward hacking and the specific power-seeking strategies employed by these AI systems \cite{amodei2016concrete, carlsmith2022powerseeking}.
\item \textbf{Aligning a Mixture of Experts (MoE) / Mixture of Agents (MoA)}. The utilization of diverse aligned models, including ensemble methods and the incorporation of additional alignment layers, remains an exciting avenue within the Mixture of Experts \cite{shen2023mixtureofexperts} and work extending this ideas into a Mixture of Agents framework \cite{wang2024mixtureofagents}.
\item \textbf{Security of Generative LLMs.} Ensuring the security of generative LLMs involves careful attention throughout the various stages of their development. Effective deployment requires a thorough grasp of the interplay and mutual influence of different methods. This encompasses the implementation of protective measures, red teaming approaches, and the combined use of techniques such as prompt engineering, fine-tuning, and retrieval-augmented generation (RAG) \cite{dong2024large, abdali2024securing}
\item \textbf{Natural Language Generation (NLG) Alignment for In-Context Learning (ICL).} Current research is delving deeper into alignment methods specifically for natural language generation \cite{lin2023unlocking}. \cite{huang2024far} surveys the state of alignment in-context learning, which underscore the lack of transparency in in-context learning (ICL) in multi-turn dialogues and instruction following, stressing the necessity for precise alignment strategies during generation to achieve more secure and dependable language models \cite{wang2024theoreticalunderstandingselfcorrectionincontext}.
\item \textbf{Natural Language Understanding (NLU) and Principled Agent.} \cite{sun2023principledriven} has investigated the formulation of models that internalize ethical principles, a promising trajectory given the potential for AI to influence discourses on AI ethics and safety \cite{Goyal_2024}. This methodology is particularly pertinent in light of the transmission of toxicity from pre-training datasets, a problem addressed by Reinforcement Learning from AI Feedback (RLAIF), also referred to as Constitutional AI \cite{bai2022constitutional}. In the RLAIF paradigm, principles are autonomously assimilated by the AI, with continuous feedback provided by the AI itself, presenting a viable approach to the scalability issues of AI systems as discussed in \cite{anwar2024foundational}. The notion of principled LLMs, as examined in recent literature, further highlights the criticality of embedding ethical considerations into the development of LLMs from their inception.
\item \textbf{Safe Reinforcement Learning for Agentic LLMs.} Reinforcement learning (RL) can be approached through a reevaluation of established concepts \cite{achiam2017constrainedpolicyoptimization, zhang2024constrainedreinforcementlearningsmoothed}, particularly unifying  RL approaches like those in control systems \cite{georgiev2024adaptive}. Furthermore, analysis of divergence within imitation learning methods also warrant review \cite{ghasemipour2019divergence}. As models gain more capability and autonomy as agents, there will be an exchange of concepts in assessing agents akin to safe RL evaluation in performing safety-critical tasks \cite{zhang2022evaluating}.
\item \textbf{Self-governing and Introspective Generative Agents.} The potential synergy between advancing natural language understanding and principled large language model research presents, along with new RL techniques, present a novel avenue for addressing alignment challenges. As AI agents can become increasingly adept at identifying vulnerabilities in their own subgoals, models equipped with self-regulation mechanisms could enhance alignment \cite{Goyal_2024}.
\end{enumerate}

Future research should rigorously investigate the confluence of AI safety and natural language generation (NLG), examining the complex interplay between language generation methodologies (e.g., next-token prediction) and the cognitive processes within AI agents. This line of inquiry has the potential to provide novel insights into addressing issues of misalignment and reward hacking as we progress further into the agentic treatment of generative LLMs.

\section{Conclusion} \label{conclusion}

In this survey paper, we explore the safety issues that accompany the accelerated research development of large language models. We examine these safety challenges that manifest at various levels: data, model, and usage. Furthermore, we delve deeply into the technical safety concerns of alignment in GAI-LLMs, emphasising the stark contrast between the proclaimed alignment criteria and the methods employed to achieve them. We shed light on the inherent safety risks that arise during the development and deployment phases of these models. Furthermore, we identify promising avenues for future research in the realm of AI safety for generative LLMs. AI alignment research is rapidly expanding into unexplored territories, encompassing a broader spectrum of techniques and challenges. From investigating safety in natural language generation models to applying knowledge distillation and navigating the complexities of multi-agent alignment, the field is brimming with opportunities for further exploration. However, the evolving landscape also introduces new complexities and trade-offs that necessitate careful consideration and rigorous investigation. The journey towards safer and more aligned AI systems demands relentless exploration and innovation, building on the foundational work presented in this comprehensive overview.

\bibliographystyle{cas-model2-names}

\bibliography{cas-refs}

\end{document}